\documentclass[aps,prd,twocolumn,showpacs,nofootinbib,floatfix]{revtex4}

\usepackage[usenames]{color}
\usepackage{amsfonts}
\usepackage{amsmath}
\usepackage{amssymb}
\usepackage{bm}
\usepackage{dcolumn}
\usepackage{epsfig}
\usepackage{graphicx}
\usepackage{graphics}
\usepackage[latin1]{inputenc}
\usepackage{latexsym}
\usepackage{rotating}
\usepackage{hyperref}
\usepackage{multirow}
\usepackage[caption=false]{subfig}
\newcommand{\<}{\begin{equation}}
\newcommand{\?}{\end{equation}}

\newcommand{\od}[2]{\left[ #1 \right]_{ #2 }}

\newcommand{\Z}{\mathbb{Z}}
\newcommand{\N}{\mathbb{N}}

\newcommand{\Q}{\mathbb{Q}}

\newcommand{\cL}{\mathcal{L}}

\DeclareMathOperator{\eulerlog}{eulerlog}

\newcommand{\ri}{\mathrm{i}}

\newcommand{\bra}{\left\langle}
\newcommand{\ket}{\right\rangle}

\begin{document}

\title{Taming the post-Newtonian expansion:\\
Simplifying
the modes of the gravitational wave energy flux at infinity
for a point particle in a circular orbit around a Schwarzschild black hole
}

\author{Nathan~K.~Johnson-McDaniel}

 \affiliation{Theoretisch-Physikalisches Institut,
  Friedrich-Schiller-Universit{\"a}t,
  Max-Wien-Platz 1,
  07743 Jena, Germany}
  
\date{\today}

%%%%%%%%%%%%%%%%%%%%%%%%%%%%%%%%%%%%%%
\begin{abstract}
High-order terms in the post-Newtonian (PN) expansions of various quantities for compact binaries exhibit a combinatorial increase in complexity, with an ever-increasing number of terms, including more transcendentals and logarithms of the velocity, higher powers of these transcendentals and logarithms, and larger and larger rational numbers as coefficients. Here we consider the gravitational wave energy flux at infinity from a point particle in a circular orbit around a Schwarzschild black hole, which is known to $22$PN [$O(v^{44})$, where $v$ is the particle's orbital velocity] beyond the lowest-order Newtonian prediction, at which point each order has over $1000$ terms. We introduce a factorization that considerably simplifies the spherical harmonic modes of the
energy flux (and thus also the amplitudes of the spherical harmonic modes of the gravitational waves); it is likely that much of the complexity this factorization removes is due to wave propagation on the Schwarzschild spacetime (e.g., tail effects).
For the modes with azimuthal number $\ell\geq 7$, this factorization reduces the expressions for the modes that enter the $22$PN total energy flux to pure integer PN series with rational coefficients, which amounts to a reduction of up to a factor of $\sim 150$ in the total number of terms in a given mode (and also in the size of the entire expression for the mode). The reduction in complexity becomes less dramatic for smaller $\ell$, due to the structure of the expansion, and one only obtains purely rational coefficients up to some order, though the factorization is still able to remove all the half-integer PN terms. For the $22$PN $\ell = 2$ modes, this factorization still reduces the total number of terms (and size) by a factor of $\sim 10$ and gives purely rational coefficients through $8$PN.
This factorization also improves the convergence of the series, though we find the exponential resummation introduced for the full energy flux in [Isoyama \emph{et al.}, Phys.\ Rev.\ D {\bf{87}}, 024010 (2013)] to be even more effective at improving the convergence of the individual modes, producing improvements of over four orders of magnitude over the original series for some modes. However, the exponential resummation is not as effective at simplifying the series, particularly for the higher-order modes.
\end{abstract}
%%%%%%%%%%%%%%%%%%%%%%%%%%%%%%%%%%%%%%


\maketitle

%%%%%%%%%%%%%%%%%%%
\section{Introduction}\label{Intro}
%%%%%%%%%%%%%%%%%%%

The standard analytic tool for study of the generation of gravitational waves from compact binaries is the post-Newtonian (PN) expansion, which has been developed to an
impressively high order over the years: For comparable mass binaries, the state of the art is $3.5$PN [$O(v^7)$ beyond Newtonian predictions, where $v$ is the binary's orbital
velocity], with some quantities known (at least partially) to higher orders. [In general, $n$PN refers to $O(v^{2n})$ past Newtonian order.] These computations are reviewed in~\cite{BlanchetLRR}.
In the extreme mass-ratio case,
where one body is much more massive than the other, and one can use black hole perturbation theory, it is far easier to calculate high-order terms in the expansion, as reviewed in~\cite{ST}. Currently, the highest-order such post-Newtonian expansion known is the $22$PN [i.e., $O(v^{44})$] expression for the gravitational energy flux and waveform at infinity due to a point
particle in a circular orbit around a Schwarzschild black hole calculated by Fujita~\cite{Fujita22PN, Fujita22PN_web}.

Like many perturbation expansions, the PN expansion exhibits a combinatorial increase of complexity at higher orders: For instance, the highest-order [$O(v^{44})$]
coefficient in the PN expansion of the Schwarzschild energy flux computed by Fujita contains $1949$ terms. These terms include logarithms of $v$, logarithms of primes up to $19$, and integer Riemann zeta function values up to $\zeta(14)$ (the ones with even arguments appear as powers of $\pi$ up to $\pi^{14}$), in addition to powers and products of all of these and the Euler-Mascheroni gamma constant $\gamma$ up to seventh order.
Moreover, many of these terms have highly complicated rational coefficients with
up to $\sim 150$ digits in both the numerator and denominator. The coefficients of the PN expansions of the individual (spin-weighted) spherical harmonic modes of the energy flux have a similarly complicated structure: The only significant difference, besides a somewhat smaller number of terms (though there are still more than $1000$ terms in the highest coefficient of some modes in the expressions needed for the $22$PN total energy flux\footnote{Since the leading-order Newtonian contribution to the higher modes is suppressed by some power of $v^2$ compared to the leading quadrupolar mode, the $22$PN expansion of the total energy flux only necessitates lower-order PN expressions for the higher modes. We shall henceforth refer to the expressions for the modes needed to obtain the total energy flux to $22$PN as the ``$22$PN energy flux expressions.''}), is that the arguments of the logarithms appearing in a given mode are restricted to $2$ and $m$,
where $m$ is the mode's degree (i.e., its magnetic quantum number).

Such complicated expressions
are obviously not easy to work with, even within a computer algebra system, and we present here a factorization that
substantially reduces this complexity. This factorization acts on the spherical harmonic modes of the energy flux (and thus also on the amplitude of the gravitational waveform), and was obtained by a combination of a study of the prime factorization of the coefficients of the energy flux, the expressions used in the black hole perturbation theory formalism of Mano, Suzuki, and Takasugi~\cite{MST1, MST2}, and the tail resummation 
introduced by Damour and Nagar~\cite{DN2007} (which also simplifies the flux modes, though not nearly as much as the factorization we introduce).

Additionally, we consider the exponential resummation proposed for the full energy flux by Isoyama~\emph{et al.}~\cite{Isoyamaetal} to improve the convergence of the series (and ensure positivity of the flux in
the Kerr case). Here we find that it considerably simplifies the individual modes (though not quite as much as the factorization, except for the highest orders), and also improves agreement with numerical calculations of the flux even more when applied to the individual modes with high degree $m$
than when applied to the full flux. We also note that all these expressions are also simplified by appropriate substitutions, including one involving the $\eulerlog_m(v)$ function introduced by Damour, Iyer, and Nagar~\cite{DIN}. 

The paper is structured as follows: We first review the relevant parts of the Mano, Suzuki, and Takasugi black hole perturbation theory formalism in Sec.~\ref{sec:MST_rev}, and then state the simplifications we consider in Sec.~\ref{sec:fac}. In Sec.~\ref{fac_disc}, we discuss how the factorizations we introduce were obtained and compare the structure of the simplifications; we illustrate the simplifications' action in Sec.~\ref{fac_ill}. We then briefly discuss how much the various factorizations and resummations we consider affect the convergence of the series in Sec.~\ref{sec:comp}, and
summarize and conclude in Sec.~\ref{sec:concl}. 
We use geometrized units throughout, with Newton's constant $G$ and the speed of light $c$ both set to unity. 

%%%%%%%%%%%%%%%%%%%
\section{The Mano, Suzuki, and Takasugi black hole perturbation theory formalism in the Schwarzschild case}
\label{sec:MST_rev}
%%%%%%%%%%%%%%%%%%%

We start by recalling a few salient facts about the computation of the energy flux and gravitational waveform at infinity using black hole perturbation theory, specifically in the
formalism of Mano, Suzuki, and Takasugi (MST)~\cite{MST1, MST2}, reviewed in~\cite{ST}. The specific expressions we use here in the Schwarzschild case (but still using the Teukolsky equation) are given in Fujita~\cite{Fujita22PN}. In this formalism, one works in the frequency domain, with gravitational wave frequency $\omega$, and also makes an expansion in (spin-weighted) spherical harmonics, indexed by $\ell$ and $m$, so $\omega = m\Omega$, where $\Omega = \sqrt{M/r_0^3}$ for a particle in a circular orbit at Schwarzschild coordinate radius $r_0$ around a black hole of mass $M$. We also have $v = \sqrt{M/r_0}$, so $M\Omega = v^3$. The basic quantities are
$Z_{\ell m\omega}$, which are computed from a solution to the Teukolsky equation. One then computes the energy flux at infinity by [Eq.~(2.11) in Fujita~\cite{Fujita22PN}]
\<\label{dEdt}
\bra\frac{dE}{dt}\ket_\infty = \sum_{\ell,m}\frac{|Z_{\ell m\omega}|^2}{4\pi\omega^2}
\?
and the waveform modes by [Eq.~(2.12) in Fujita~\cite{Fujita22PN}]
\<\label{hlm}
h_{\ell m} = -\frac{2}{r}\frac{Z_{\ell m\omega}}{\omega^2}e^{\ri\omega(r^*-t)},
\?
where $r^* = r + 2M\log(r/2M - 1)$ is the tortoise coordinate and we have chosen our coordinate system to remove a constant phase shift [cf.\ Eq.~(4.3b) in Fujita and Iyer~\cite{FI}].

We now discuss how one computes $Z_{\ell m\omega}$. Fundamental to the
MST approach is the introduction of the renormalized angular momentum $\nu$, which allows the solution to the Teukolsky (or Regge-Wheeler) equation to be expressed as a series of Coulomb wave
functions, following a line of ideas first introduced in general relativity by Leaver~\cite{LeaverJMP} (and dating back to relatively early work on the two-center problem in quantum mechanics, as discussed by Leaver). One fixes $\nu$
by demanding that the series converge and finds that it is given by the solution of a continued fraction equation [Eq.~(3.5) in Fujita~\cite{Fujita22PN}; Eq.~(3.12) is used in practice]. Specifically, it is the solution to that equation that reduces to $\ell$ when $v\to 0$. [\emph{Nota bene} (N.B.): $\nu$ depends upon $\ell$, $m$, and $v$, though it is not customary in the literature to indicate any of this dependence explicitly.] If one performs a post-Newtonian expansion of $\nu$
for Schwarzschild, one finds that it has the form
\<
\nu = \ell + \sum_{k = 1}^\infty \od{\nu_\ell}{k} (2 m v^3)^{2k},
\?
where $\od{\nu_\ell}{k}\in\Q$. Bini and Damour give $\od{\nu_\ell}{k}$ for $k\in\{1,2,3\}$ in the Appendix of~\cite{BD2} [where it is referred to as $\nu_{2k}(\ell)$], though we need $\nu$ to considerably higher orders than given by Bini and Damour [to $O(v^{42})$, i.e., $k = 7$, for $\ell \in \{2,3\}$]. Expansions to $O(v^{54})$ have been calculated for us by Abhay G.\ Shah and are included in the electronic material accompanying this article~\cite{data_URL}.

One then obtains $Z_{\ell m\omega}$ from
\<\label{Zlmo}
Z_{\ell m\omega} = \frac{\cL_{\ell m\omega} R^\text{in}_{\ell m\omega}}{B^\text{inc}_{\ell m\omega}},
\?
where $\cL_{\ell m\omega}$ is a linear, second-order differential operator (which differentiates with respect to $r_0$, the Schwarzschild coordinate orbital radius) given in
Eq.~(2.9) in Fujita~\cite{Fujita22PN}, and [Eqs. (3.8), (3.10b), and (3.11a) in Fujita~\cite{Fujita22PN}]
\begin{subequations}
\begin{align}
\label{Rin}
R^\text{in}_{\ell m\omega} &= K_\nu R^\nu_\text{C} + K_{-\nu-1}R^{-\nu-1}_\text{C},\\
\label{Binc}
B^\text{inc}_{\ell m\omega} &= \frac{1}{\omega}\left[K_\nu - \ri e^{-\ri\pi\nu}\frac{\sin\pi(\nu+\ri\epsilon)}{\sin\pi(\nu-\ri\epsilon)}K_{-\nu-1}\right]A_+^\nu\epsilon^{-\ri\epsilon},\\
\label{A+}
A_+^\nu &= 2^{-3-\ri\epsilon}e^{-\pi\epsilon/2}e^{\ri\pi(\nu+3)/2}\frac{\Gamma(\nu + 3 + \ri\epsilon)}{\Gamma(\nu - 1 - \ri\epsilon)}\sum_{n=-\infty}^{\infty}a_n^\nu,
\end{align}
\end{subequations}
where $\epsilon := 2M\omega = 2m v^3$, $\Gamma$ is the gamma function,  and $a_n^\nu$ are the coefficients in the Coulomb wave function expansion of $R_\text{C}^\nu$; see Eq.~\eqref{RC}, below. [These coefficients are given explicitly for $-2\leq n \leq 2$ through $O(\epsilon^2)$ in Eqs.~(3.15) in Fujita and Iyer~\cite{FI} (note that $n_0^\nu = 1$).] We also have [Eq.~(3.9) in Fujita~\cite{Fujita22PN}]
\begin{widetext}
\<
\label{Knu}
\begin{split}
K_\nu &= \frac{2^2\ri^Ne^{\ri\epsilon}(2\epsilon)^{-2-\nu-N}\Gamma(3-2\ri\epsilon)\Gamma(N+2\nu+2)}{\Gamma(N+\nu+3+\ri\epsilon)\Gamma(N+\nu+1+\ri\epsilon)\Gamma(N+\nu-1+\ri\epsilon)}\\
&\quad \times\left[\sum_{n=N}^\infty(-1)^n\frac{\Gamma(n+N+2\nu + 1)}{(n-N)!}\frac{\Gamma(n+\nu-1+\ri\epsilon)\Gamma(n+\nu+1+\ri\epsilon)}{\Gamma(n+\nu+3-\ri\epsilon)\Gamma(n+\nu+1-\ri\epsilon)}a_n^\nu\right]\\
&\quad \times\left[\sum_{n=-\infty}^N\frac{(-1)^n}{(N-n)!(N+2\nu+2)_n}\frac{(\nu-1-\ri\epsilon)_n}{(\nu+3+\ri\epsilon)_n}a_n^\nu\right]^{-1},
\end{split}
\?
where $N\in\Z$ is arbitrary ($K_\nu$ is independent of $N$, despite appearances) and $(x)_n := \Gamma(x+n)/\Gamma(x)$ is the Pochhammer symbol. Finally, we have [Eqs. (3.1) and (3.2) in
Fujita~\cite{Fujita22PN}, evaluated at $r = r_0$]
\<
\label{RC}
R_\text{C}^\nu = (\omega r_0)^2\left(1 - \frac{\epsilon}{\omega r_0}\right)^{2-\ri\epsilon}e^{-\ri \omega r_0}\sum_{n=-\infty}^\infty(-\ri)^n(2\omega r_0)^{n+\nu}(\nu-1-\ri\epsilon)_n\frac{\Gamma(\nu + 3 + \ri\epsilon)}{\Gamma(2n+2\nu + 2)}a_n^\nu\Phi(n+\nu + 3 + \ri\epsilon, 2n + 2\nu + 2; 2\ri\omega r_0),
\?
\end{widetext}
where $\Phi(\alpha, \beta; z)$ is the confluent hypergeometric function that is regular at $z = 0$.

%%%%%%%%%%%%%%%%%%%%%%%
\section{Simplifying the modes of the energy flux}
\label{sec:fac}
%%%%%%%%%%%%%%%%%%%%%%%

Here we consider seven different ways to simplify the (spin-weighted) spherical harmonic modes of the gravitational wave energy flux from a point particle in a circular orbit about a Schwarzschild black hole: Performing a substitution, computing the PN expansion of the logarithm of the flux, and four different factorizations, as well as combining the expansion of the logarithm with the most effective factorization. In all cases, the substitutions act on the individual (spin-weighted spherical harmonic) modes of the flux $\eta_{\ell m}$, defined so that the total energy flux is given by
\<
\bra\frac{dE}{dt}\ket_\infty =: \bra\frac{dE}{dt}\ket_\text{Newt}\sum_{\ell, m}\eta_{\ell m},
\?
where $\bra dE/dt\ket_\text{Newt} = (32/5)(\mu/M)^2v^{10}$, $\mu$ is the reduced mass, and the sum is taken only over $m\geq 1$ (there are no $m = 0$ contributions, and the negative $m$ values just would duplicate the positive ones, so they are lumped together, for convenience). We mostly consider $\bar{\eta}_{\ell m}$, defined to be $\eta_{\ell m}$ with the lowest-order piece factored out, so the first term in the series is $1$ (so $\bar{\eta}_{22} = \eta_{22}$, but all the other modes are modified). Note that the leading power of $v$ in $\eta_{\ell m}$ is given by
\<
p_{\ell m} =
\begin{cases}
2\ell - 4 &\text{if $\ell + m$ is even},\\
2\ell - 2 &\text{if $\ell + m$ is odd},
\end{cases}
\?
so that $\bar{\eta}_{\ell m}$ is known to $(22 - p_{\ell m}/2)$PN for the $22$PN total energy flux.
The leading coefficient (which is rational) can be deduced from the expressions for the leading term in the PN expansion of the waveform modes in Eqs.~(330) in Blanchet's Living Review~\cite{BlanchetLRR}; explicit expressions are given in the {\sc{Mathematica}} notebook associated with this paper~\cite{data_URL}. Note also that the waveform modes are given  to $2.5$PN for a general $\ell$ and $m$ by Fujita and Iyer~\cite{FI}, though they leave their expressions in terms of spin-weighted spherical harmonics, while the expressions obtain from Blanchet are completely explicit. 

The simplifying substitution is given using the function
\<
\eulerlog_m(v) := \gamma + \log(2mv),
\?
first introduced by Damour, Iyer, and Nagar~\cite{DIN}. Here $\gamma$ denotes the Euler-Mascheroni gamma constant. We have modified their definition slightly since we are using a different expansion parameter; cf.\ Fujita and Iyer~\cite{FI}, who refer to this as $\eulerlog(m,v)$. We also find that the logarithms remaining after the $\eulerlog_m(v)$ substitution can all be written in terms of $\log(2v^2)$. Specifically, one performs the substitutions
\begin{subequations}
\begin{align}
\gamma &\to \eulerlog_m(v) - \log 2 - \log m  - \log v ,\\
\log 2 &\to \log(2v^2) - 2\log v
\end{align}
\end{subequations}
(in this order).

Another way of simplifying these results is related to the work of Isoyama \emph{et al.}~\cite{Isoyamaetal}, who advocated PN expanding $\log\bra dE/dr\ket_\infty$ and then exponentiating (with no expansion) as a way to
improve the convergence of the series, and ensure its positivity in the Kerr case. Here we note that PN expanding $\log\bar{\eta}_{\ell m}$ 
produces a considerable simplification of its analytic form.

We also consider the leading logarithm tail factorization introduced by Damour and Nagar~\cite{DN2007}, which is given by $\bar{\eta}_{\ell m}/|T_{\ell m}|^2$, where 
\<\label{Tlm}
|T_{\ell m}| = e^{\pi m v^3}\frac{|\Gamma(1+\ell - 2\ri m v^3)|}{\Gamma(1+\ell)}.
\?
The full version of the factorization [given in, e.g., Eq.~(16) of Damour and Nagar~\cite{DN2007}] also includes a phase factor that is not relevant to the present discussion and was introduced with slightly different notation---$T_{\ell m}$ is now common in the literature.
Note that we can use the standard gamma function identity (from the reflection formula) $|\Gamma(1+\ri z)|^2 = \pi z/\sinh\pi z$  [(6.1.31) in Abramowitz and Stegun~\cite{AS}] along with the
gamma function's recurrence relation to write
\<\label{Tlm2}
|T_{\ell m}|^2 = \frac{4\pi m v^3}{1 - e^{- 4\pi mv^3}}\prod_{k = 1}^\ell\left[1 + \left(\frac{2mv^3}{k}\right)^2\right]
\?
[cf.\ Eq.~(59) of Damour, Iyer, and Nagar~\cite{DIN}, which presents this expression in a slightly different form]. Additionally, note that the leading logarithms that $T_{\ell m}$ resums occur in the phase factor and are thus not shown here.

%..................
\subsection{The $S_{\ell m}$ factorization}
%.................

The first simplifying factorization we introduce is given using
\<
\label{Slm}
S_{\ell m} := (2m v)^{\bar{\nu}_{\ell m}(v)}e^{\pi m v^3}\frac{\Gamma[1 + \bar{\nu}_{\ell m}(v) - 2\ri m v^3]}{\Gamma[1+2\bar{\nu}_{\ell m}(v)]},
\?
where $\bar{\nu}_{\ell m}(v) := \nu - \ell$. [We show the dependence of $\bar{\nu}_{\ell m}(v)$ on $\ell$, $m$, and $v$ explicitly, for clarity, even though it is not customary in the literature to do this for $\nu$.]

Specifically, we have
\<
\frac{\bar{\eta}_{\ell m}}{|S_{\ell m}|^2} = 1 + \sum_{k = 1}^{3+2\ell}\alpha_kv^{2k} + O(v^{8+4\ell}),
\?
where $\alpha_k\in\Q$. Thus, through $O(v^{7 + 4\ell})$, factoring out $|S_{\ell m}|^2$ removes all the transcendentals, as well as the odd powers and logarithms of $v$,
and leaves a pure integer order PN series with rational coefficients. (N.B.: These expressions
contain the Euler-Mascheroni gamma constant  $\gamma$ and the Riemann zeta function evaluated at odd integers. These numbers are not known to be transcendental, or in
many cases even irrational. However, they are all strongly conjectured to be transcendental, so we shall refer to them all as such.) This means that for $\ell\geq 7$, this factorization turns the $22$PN total energy flux results for $\eta_{\ell m}$ into such purely rational integer order
PN series. Moreover, even higher-order terms that still contain
transcendentals and $\log v$ terms are significantly simplified by this
factorization, as illustrated in Sec.~\ref{fac_ill}. One obtains the same simplification upon factoring out $|S_{\ell m}|$ from $|h_{\ell m}|$ (i.e., the amplitude of the
gravitational wave modes), as one would expect from Eqs.~\eqref{dEdt} and~\eqref{hlm}.

We can write $S_{\ell m}$ in a form that better illustrates some of its structure, and makes for faster computations in {\sc{Mathematica}} using the expansion
\<\label{gamma-seq}
\Gamma(1 + z) = \exp\left[-\gamma z + \sum_{n=2}^\infty\frac{\zeta(n)}{n}(-z)^n\right],
\?
which gives
\begin{subequations}
\label{Slm_alt}
\<
S_{\ell m} = \exp\left[\bar{\nu}_{\ell m}(v)\eulerlog_m(v) + \pi mv^3 + \sigma_{S_{\ell m}}(v)\right],
\?
\<
\sigma_{S_{\ell m}}(v) := \sum_{n = 2}^\infty\frac{\zeta(n)}{n}\bigl\{\left[-\bar{\nu}_{\ell m}(v) + 2\ri m v^3\right]^n - [-2\bar{\nu}_{\ell m}(v)]^n\bigr\}.
\?
\end{subequations}

%..................
\subsection{The $S_{\ell m}V_{\ell m}$ factorization}
%.................

\begin{table*}
\caption{\label{coeff_table} The values of $-\od{\nu_\ell}{1}$ [$-\nu_2(\ell)$ in Bini and Damour~\cite{BD2}], $\bar{q}_\ell$, $-(\bar{q}_\ell\od{\nu_\ell}{1})^{-1}$, $s_\ell$, and $-(s_\ell\od{\nu_\ell}{1})^{-1}$ for $\ell\leq 6$. For $\ell\in\{5,6\}$ we do not give values for the last two quantities, since they cannot be determined from the $22$PN energy flux expressions. We give the prime factorizations of $-\od{\nu_\ell}{1}$,
$-(\bar{q}_\ell\od{\nu_\ell}{1})^{-1}$, and $-(s_\ell\od{\nu_\ell}{1})^{-1}$ in order to illustrate their structure.
}
\begin{tabular}{*{6}{c}}
\hline\hline
$\ell$ & $-\od{\nu_\ell}{1}$ & $\bar{q}_\ell$ & $-(\bar{q}_\ell\od{\nu_\ell}{1})^{-1}$ & $s_\ell$ & $-(s_\ell\od{\nu_\ell}{1})^{-1}$\\
\hline
2 & $\frac{107}{210} = \frac{107^1}{2^13^15^17^1}\phantom{X}\,$ & $\frac{7}{214}$ & $2^23^15^1$ & $\frac{7}{17120}$ & $2^63^15^2$\\
3 & $\frac{13}{42} = \frac{13^1}{2^13^17^1}\phantom{X}\;\;$ & $\frac{1}{520}$ & $2^43^15^17^1$ & $\frac{1}{10483200}$ & $2^{10}3^35^27^2$\\
4 & $\frac{1571}{6930} = \frac{1571^1}{2^13^25^17^111^1}\;$ & $\frac{11}{87976}$ & $2^43^25^17^2$ & $\frac{11}{595928309760}$ & $2^{14}3^55^27^4$\\
5 & $\frac{773}{4290} = \frac{773^1}{2^13^15^111^113^1}$ & $\frac{13}{1558368}$ & $2^63^35^17^111^1$ & -- & --\\
6 & $\frac{901}{6006} = \frac{17^153^1}{2^13^17^111^113^1}$ & $\frac{1}{1783980}$ & $2^33^35^17^111^213^1$ & -- & --\\
\hline\hline
\end{tabular}
\end{table*}

One can remove some more transcendentals and logarithms from $\bar{\eta}_{\ell m}/|S_{\ell m}|^2$ by additionally factoring out $|V_{\ell m}|^2$, where 
\<
\label{Vlm}
V_{\ell m} := \frac{V_{\ell m}^\text{num}}{V_{\ell m}^\text{denom}},
\?
\begin{widetext}
\begin{subequations}
\label{Vlm_parts}
\<
\begin{split}
V_{\ell m}^\text{num} &:= 1 + q_{\ell m}(2v^2)^{1+2\ell+2\bar{\nu}_{\ell m}(v)}\frac{\Gamma[1-2\bar{\nu}_{\ell m}(v)]}{\Gamma[1+2\bar{\nu}_{\ell m}(v)]}
\left\{\frac{\Gamma[1 + \bar{\nu}_{\ell m}(v) - 2\ri mv^3]}{\Gamma[1 - \bar{\nu}_{\ell m}(v) - 2\ri mv^3]}\right\}^2\\
&\,= 1 + q_{\ell m}(2v^2)^{1+2\ell}\exp\left[2\bar{\nu}_{\ell m}(v)\log(2v^2) + \sigma_{V^\text{num}_{\ell m}}(v)\right],
\end{split}
\?
\<
\begin{split}
V_{\ell m}^\text{denom} &:= 1 + \ri s_\ell(4mv^3)^{1+2\ell+2\bar{\nu}_{\ell m}(v)}e^{-\ri\pi\bar{\nu}_{\ell m}(v)}
\frac{\bar{\nu}_{\ell m}(v) + 2\ri mv^3}{\bar{\nu}_{\ell m}(v) - 2\ri mv^3}
\left\{\frac{\Gamma[1 - 2\bar{\nu}_{\ell m}(v)]}{\Gamma[1 + 2\bar{\nu}_{\ell m}(v)]}\right\}^2\left\{\frac{\Gamma[1 + \bar{\nu}_{\ell m}(v) - 2\ri mv^3]}{\Gamma[1 - \bar{\nu}_{\ell m}(v) - 2\ri mv^3]}\right\}^3\\
&\,= 1 + \ri s_\ell(4mv^3)^{1+2\ell}\frac{\bar{\nu}_{\ell m}(v) + 2\ri mv^3}{\bar{\nu}_{\ell m}(v) - 2\ri mv^3}\exp\left\{\bar{\nu}_{\ell m}(v)[2\eulerlog_m(v) +2 \log(2v^2) - \ri\pi] + \sigma_{V^\text{denom}_{\ell m}}(v)\right\},
\end{split}
\?
\<
\begin{split}
\sigma_{V^\bullet_{\ell m}}(v) := \sum_{n = 2}^\infty\frac{\zeta(n)}{n}\left(A_\bullet\left\{[2\bar{\nu}_{\ell m}(v)]^n - [-2\bar{\nu}_{\ell m}(v)]^n\right\} + B_\bullet\bigl\{\left[-\bar{\nu}_{\ell m}(v) + 2\ri m v^3\right]^n - \left[\bar{\nu}_{\ell m}(v) + 2\ri m v^3\right]^n\bigr\}\right),\\
[A_\text{num} = 1,\quad B_\text{num} = 2;\quad A_\text{denom} = 2,\quad B_\text{denom} = 3],
\end{split}
\?
\end{subequations}
\end{widetext}
with $q_{\ell m}, s_\ell\in\Q$ constants that are determined by requiring that the factorization removes certain terms. The values for these constants that it is possible to determine from the $22$PN energy flux
expressions for the modes are given
in Table~\ref{coeff_table}, where we write $q_{\ell m} = \beta_{\ell m}\bar{q}_\ell$, with
\<
\beta_{\ell m} :=
\begin{cases}
1& \text{if $\ell + m$ is even,}\\
-\frac{\ell + 1}{\ell}& \text{if $\ell + m$ is odd}.
\end{cases}
\?
We have also given alternative forms of $V^\text{num, denom}_{\ell m}$ in terms of $\eulerlog_m(v)$ and $\log(2v^2)$ that display their structure somewhat differently (and that we actually use for computing their expansions).

We obtain the constants $q_{\ell m}$ and $s_\ell$ by demanding that  factoring out $|V_{\ell m}|^2$ from $\bar{\eta}_{\ell m}$ removes the $\log(2v^2)v^{8+4\ell}$ and the $v^{9+6\ell}$ term, respectively. We thus are unable to determine $q_{\ell m}$ for $\ell\geq 7$ and $s_\ell$ for $\ell\geq 5$
from the $22$PN energy flux expressions: For $q_{77}$, we would need to know the $v^{36}$ term in $\bar{\eta}_{77}$, but only know this through $v^{34}$. Similarly, for $s_5$, we would need to know, e.g., the $v^{39}$ term of $\bar{\eta}_{55}$ but only know this through $v^{38}$. Thus, while it appears that $q_{\ell m}$ and $s_\ell$ are simply related to $1/\od{\nu_\ell}{1}$, as illustrated in Table~\ref{coeff_table}, we do not know them for sufficiently many values of $\ell$ to be able to deduce the specific relation with any confidence. 

The additional simplification from factoring out $|V_{\ell m}|^2$ from $\bar{\eta}_{\ell m}/|S_{\ell m}|^2$ is not nearly as dramatic as that from factoring out $|S_{\ell m}|^2$ from $\bar{\eta}_{\ell m}$.  Nevertheless, it is possible that a slightly different combination of gamma functions in $V_{\ell m}$ could remove further terms, since there is still a fair amount of structure in the remaining transcendentals, as is illustrated in Sec.~\ref{fac_ill}.

%..................
\subsection{The $S_{\ell m}V'_{\ell m}$ factorization}
\label{ssec:Vplm}
%.................

Moreover, one can remove the remaining odd powers of $v$ by making the substitution
\<\label{s-series}
s_\ell \to s_\ell\left[1 + \sum_{k=1}^\infty \od{\bar{s}_{\ell}}{k}(2mv^3)^{2k}\right],
\?
in $V_{\ell m}$, where one fixes $\od{\bar{s}_{\ell}}{k}$ by demanding that the factorization remove the $v^{9+6(\ell + k)}$ term from $\bar{\eta}_{\ell m}$. We will use $V'_{\ell m}$ to denote $V_{\ell m}$ with the substitution in
Eq.~\eqref{s-series}. Here one can only fix the lowest few of these coefficients using the $22$PN energy flux expressions, obtaining
\begin{subequations}
\label{s2}
\begin{align}
\od{\bar{s}_2}{1} &= \frac{416607433}{56624400}+\frac{1 }{3}\pi ^2,\\
\od{\bar{s}_2}{2} &= \frac{46804742792313761}{1469564559540000}+\frac{15520
   3051}{56624400} \pi ^2-\frac{1 }{45}\pi ^4,\\
\od{\bar{s}_2}{3} &= \frac{19229488138491297298603997}{180295291464
   636348000000}\nonumber\\
   &\quad +\frac{59240111985731}{4535693085000}\pi
   ^2-\frac{3297719}{849366000} \pi
   ^4+\frac{2}{945}\pi ^6\nonumber\\
   &\quad -\frac{28
   }{1605}[\eulerlog_m(v) +\log(2v^2)],
\end{align}
\end{subequations}
\begin{subequations}
\begin{align}
\od{\bar{s}_3}{1} &= \frac{72823147}{10810800}+\frac{1 }{3}\pi ^2,\\
\od{\bar{s}_3}{2} &=  \frac{2314965899122031}{95446607256000}+\frac{177976343 }{75675600}\pi
   ^2-\frac{1 }{45}\pi ^4,
\end{align}
\end{subequations}
\<
\od{\bar{s}_4}{1} = \frac{37602861148067}{5884875196200}+\frac{1}{3} \pi ^2.
\?
Note that $\od{\bar{s}_2}{3}$ depends on $m$, though we do not denote this explicitly, since the dependence only arises through $\eulerlog_m(v)$. Also note that the highest power of $\pi$ in $\od{\bar{s}_{\ell}}{k}$ seems to be given by $2(-1)^{k+1}\zeta(2k)$, independent of $\ell$, since $\zeta(2) = \pi^2/6$, $\zeta(4) = \pi^4/90$, and $\zeta(6) = \pi^6/945$. Thus, one may conjecture that these come from some sort of (poly)gamma function expansion.

We also note that the series by which one replaces $s_\ell$ in $V'_{\ell m}$ [i.e., Eq.~\eqref{s-series}] is also somewhat simpler when one expands its logarithm, though here the simplification is quite mild: One merely obtains nicer-looking prime factorizations for some of the coefficients at higher orders in the case with the logarithm. For instance, for $\od{\bar{s}_2}{2}$, the coefficient of $\pi^2$ changes from
$155203051^1/2^43^35^27^2107^1$ to $107^2/2^13^45^17^2$ upon taking the logarithm of the series, with similar simplifications for the coefficient of $\pi^2$ in $\od{\bar{s}_3}{2}$ ($177976343^1/2^43^35^27^211^113^1 \to 5^113^2/2^13^47^2$) and the coefficient of $\pi^4$
in $\od{\bar{s}_2}{3}$ ($-211^115629^1/2^43^45^37^2107^1 \to -11^1107^2/2^23^45^37^2$); cf.\ the values of $\od{\nu_\ell}{1}$ given in Table~\ref{coeff_table}. The coefficient of $\pi^2$ in $\od{\bar{s}_2}{3}$ is also somewhat simplified, though the simplification is not so dramatic ($179^1743^1445424423^1/2^33^15^47^411^1107^2 \to 89^1107^1888011^1/2^33^75^47^4$). Note, however, that the coefficients of the highest power
of $\pi$ are somewhat less simple after taking the logarithm (e.g., $2^1/3^35^17^1 \to 2^131^1/3^45^17^1$ in $\od{\bar{s}_2}{3}$).

It is possible to make a similar substitution for $q_{\ell m}$, taking
\<\label{qlm_subs}
q_{\ell m} \to q_{\ell m}\left[1 + \sum_{k=1}^\infty\od{\bar{q}_{\ell m}}{k}v^{2k}\right]
\?
and fixing $\od{\bar{q}_{\ell m}}{k}\in\Q$ by demanding that the factorization remove the $\log(2v^2)$ term in the coefficient of $v^{8+4\ell + 2k}$ in $\bar{\eta}_{\ell m}$. However, this is not as efficacious as the similar substitution for $s_{\ell}$: Indeed, in general the $V_{\ell m}^\text{denom}$ 
factorization removes far more terms than the $V_{\ell m}^\text{num}$ factorization. For instance, for $\eta_{22}$, the $V^\text{num}_{\ell m}$ factorization removes $18$ terms (in $4$ coefficients) through $22$PN, while the $V^\text{denom}_{\ell m}$ factorization removes $42$ terms through that order with just the single rational value for $s_\ell$, and $123$ terms when one uses the series given in Eqs.~\eqref{s2}. In both cases the $V^\text{denom}_{\ell m}$ factorization removes terms from $11$ coefficients, in the latter case setting them all to zero.

Here we find that fixing $\od{\bar{q}_{\ell m}}{k}$ as described above only removes terms in the coefficients of $v^{8+4\ell + 2k + 6n}$, $n\in\N_0$,\footnote{We use $\N_0$ to denote the positive integers including zero, reserving $\N$ for the strictly positive integers.} and even then does
not remove all the transcendentals and terms involving $\log v$. In fact, for $k > \ell - 2$, it only removes one term at each order [the highest power of $\log(2v^2)$], though it does slightly simplify the prime factorizations of the coefficients of some of the remaining terms.
For $k \leq \ell - 2$, the simplification is more substantial: 
For example, factoring $V_{\ell m}$ out of $\bar{\eta}_{33}$ and including $\od{\bar{q}_{33}}{1} = 7$ removes the $\log^{1+ n}(2v^2)v^{22+6n}$, $\zeta(3+2n)v^{28+6n}$, and $\zeta(3+2n)\log^{1+n}(2v^2)v^{34+6n}$ terms ($n\in\N_0$), at least through $O(v^{40})$, the highest relevant order of the $22$PN energy flux results, as well as the $\zeta^2(3)$ term in the coefficient of $v^{40}$ itself. Moreover, it converts the
numerator of the coefficient of $\pi^4 v^{34}$ from $2^{10}3^213^137^1$ to $2^{11}3^213^1$ (leaving the denominator unchanged), and produces a similar simplification of the numerator of the coefficient of $\pi^6v^{40}$ (viz., $2^{13}3^413^117^1 \to 2^{14}3^313^1$).
We will not consider this substitution further here, except to note that the $\od{\bar{q}_{\ell m}}{k}$ for the same $\ell$ do not appear
to be related simply, unlike the original $q_{\ell m}$ (and that the values of $\od{\bar{q}_{\ell m}}{k}$ are generally more complicated rationals than $\od{\bar{q}_{33}}{1}$ is).

%%%%%%%%%%%%%%%%%%%%%
\section{Discussion of the factorizations}
\label{fac_disc}
%%%%%%%%%%%%%%%%%%%%%

One can see how these factorizations arise in the MST formalism by noting that $\eta_{\ell m} \propto |Z_{\ell m\omega}|^2$ [cf.\ Eq.~\eqref{dEdt}], while
\<\label{Zlmop}
\begin{split}
Z_{\ell m\omega} \sim \frac{R^\text{in}_{\ell m\omega}}{B^\text{inc}_{\ell m\omega}} &\sim \frac{R^\nu_\text{C}}{A_+^\nu}\left[1 + \frac{K_{-\nu-1}}{K_\nu}\frac{R^{-\nu-1}_\text{C}}{R^\nu_\text{C}}\right]\\
&\quad\times \left[1 - \ri e^{-\ri\pi\nu}\frac{\sin\pi(\nu + \ri\epsilon)}{\sin\pi(\nu - \ri\epsilon)}\frac{K_{-\nu-1}}{K_\nu}\right]^{-1}
\end{split}
\?
from Eqs.~\eqref{Zlmo}, \eqref{Rin}, and \eqref{Binc}, where the use of $\sim$ indicates that we have neglected contributions that will not contribute transcendentals or $\log v$ terms, including the action of the linear operator $\cL_{\ell m\omega}$ on $R^\text{in}_{\ell m\omega}$.
Additionally, we have
\begin{subequations}
\begin{align}
|R^\nu_\text{C}| &\sim (2mv)^\nu\frac{|\Gamma(1+\nu + \ri\epsilon)|}{\Gamma(1+2\nu)},\\
|A^\nu_+| &\sim e^{-\pi\epsilon/2},\\
\frac{K_{-\nu-1}}{K_\nu} &\sim (2\epsilon)^{1+2\nu}\left[\frac{\Gamma(1-2\nu)\Gamma(1+\nu-\ri\epsilon)}{\Gamma(1+2\nu)\Gamma(1-\nu-\ri\epsilon)}\right]^2\frac{\Gamma(1+\nu+\ri\epsilon)}{\Gamma(1-\nu+\ri\epsilon)},
\end{align}
\end{subequations}
where we have used the circular orbit expression $\omega r_0 = mv$ and recalled that $\Gamma(n+x)=\Gamma(1+x)\prod_{k=1}^{n-1}(k+x)$ for $n\in\N$, $n\geq 2$ (from the gamma function's recurrence relation), so we can replace any integer in the argument of a gamma function by $1$ without changing the transcendental
content of the expression. In particular, this lets us factor out the gamma functions from inside sums.

Note that the expansion of the confluent hypergeometric function in $R_\text{C}^\nu$ does not generate any transcendentals, even though one expands in all its arguments: Since its final argument is proportional to $v$, only a finite number of rationals contribute at a given PN order, so no transcendentals are generated. Similarly, the expansion of the Pochhammer symbol does not lead to transcendentals, since the expression in terms of gamma functions is only a convenient way of representing it: $(x)_n$ is a rational function of $x$ for $n\in\Z$.

N.B.: While the $V_{\ell m}$ factorization was obtained from an inspection of the pieces that contribute to the calculation in the MST formalism, the $S_{\ell m}$ factorization was mostly obtained by
a study of the structure of the prime factorization of the coefficients of the expansion, combined with the $T_{\ell m}$ factorization. The intuition gained from comparing this result with the MST formalism then allowed us to conjecture the $V_{\ell m}$ factorization from the structure of the MST formalism.

We thus see that $|S_{\ell m}|$ comes from $|R^\nu_\text{C}/A^\nu_+|$, while $V_{\ell m}^\text{num}$ and $V_{\ell m}^\text{denom}$ correspond to the first and second term in brackets in Eq.~\eqref{Zlmop}, respectively. [N.B.: We have simplified
the expression for $V_{\ell m}^\text{denom}$ by using the relation $\Gamma(1+z) \sin\pi z = \pi z/\Gamma(1-z)$, which comes from the reflection formula, (6.1.17) in Abramowitz and Stegun~\cite{AS}.] The fact that the $V_{\ell m}$ 
contributions only arise at relatively high orders is a consequence of the fact that the $K_{-\nu-1}$ contributions only enter at such high orders,
as discussed around Eq.~(6.10) in~\cite{MST2} for the Regge-Wheeler version of the MST formalism.

Comparing the $S_{\ell m}$ factorization [Eq.~\eqref{Slm}] to the Damour-Nagar leading logarithm tail factorization that uses $T_{\ell m}$ [whose amplitude is given in Eq.~\eqref{Tlm}], we see that they are indeed very similar. 
In fact, $|S_{\ell m}|$ only differs from $|T_{\ell m}|$ by the replacement of
$\ell$ by $\bar{\nu}_{\ell m}(v)$, the factor of $2$ multiplying $\bar{\nu}_{\ell m}$ in the gamma function in the denominator, and the $(2 m v)^{\bar{\nu}_{\ell m}(v)}$ factor, though all of these differences are very important to the transcendentality structure of the PN expansion of $S_{\ell m}$. Indeed, it is clear from the expression for $|T_{\ell m}|^2$ given in Eq.~\eqref{Tlm2} that the $T_{\ell m}$ factorization
cannot remove nearly so many transcendentals as the $S_{\ell m}$ factorization (only the powers of $\pi$, and not even all of those), as seen in Sec.~\ref{fac_ill}.
In $S_{\ell m}$, one gets $\zeta(n)$ terms from the expansion of the gamma functions, since $\bar{\nu}_{\ell m}(v)$ is not an integer. One also gets $\gamma$ terms from the gamma
functions in $S_{\ell m}$ due to the presence of a different dependence on $\bar{\nu}_{\ell m}(v)$ in the argument of the gamma functions in the numerator and denominator.
The alternate expressions we give for $S_{\ell m}$ and $V_{\ell m}$ in terms of $\eulerlog_m(v)$ and $\log(2v^2)$ [in Eqs.~\eqref{Slm_alt} and~\eqref{Vlm_parts}] display the structure of the expansions more explicitly and help explain why the $\eulerlog_m(v)$ and $\log(2v^2)$ substitutions and expanding the logarithm both produce a simplification. 

Note also that Fujita comments on how some portions of the MST formalism generate the transcendentals appearing in the PN expansion of the energy flux in Appendix A of~\cite{Fujita22PN}. Fujita additionally  remarks that some of the
logarithms for $\ell = 2$ can be explained by the renormalization group arguments from Goldberger~\emph{et al.}~\cite{GR, GRR}, who derive them using the beta function for the mass quadrupole moment, 
which describes how its value changes with scale. Physically, this scale dependence is due to tail effects, i.e., backscattering of the radiation off of the background curvature, which are discussed in, e.g.,~\cite{BlanchetLRR}. (See~\cite{FStail} for further discussion of the renormalization group approach to tail effects.) Indeed, the factor of $\od{\nu_2}{1} = 107/210$ is also found in the second-order tail effects calculated by Blanchet~\cite{Blanchet1998} in the form of $1712/105 = 2^5\od{\nu_2}{1}$; see, e.g., Eq.~(5.9) in Blanchet~\cite{Blanchet1998}. Additionally, it is possible that more of $S_{\ell m}$ could be obtained by these effective field theory methods using the matching procedure mentioned in Goldberger and Ross~\cite{GR}.

It is also worth noting that the Damour-Nagar $T_{\ell m}$ factorization resums tail effects, and can be derived from a Coulomb wave function calculation (thus showing the similarities to the MST formalism), following Asada and Futamase~\cite{AF}. (See also Khriplovich and Pomeransky~\cite{KP} for further intuition about higher-order tail effects.) In fact, it seems possible that in general the renormalized angular momentum, $\nu$, is related to
the tail effects (to all orders) and thus also to the beta functions for the multipole moments. The $\eulerlog_m(v)$ function that appears in these results is also related to the tails---cf.\ Eqs.~(310) in Blanchet's Living Review~\cite{BlanchetLRR}.

%%%%%%%%%%%%%%%%%%%%%%
\section{Illustration of the simplifications}
\label{fac_ill}
%%%%%%%%%%%%%%%%%%%%%%

\begin{figure*}[htb]
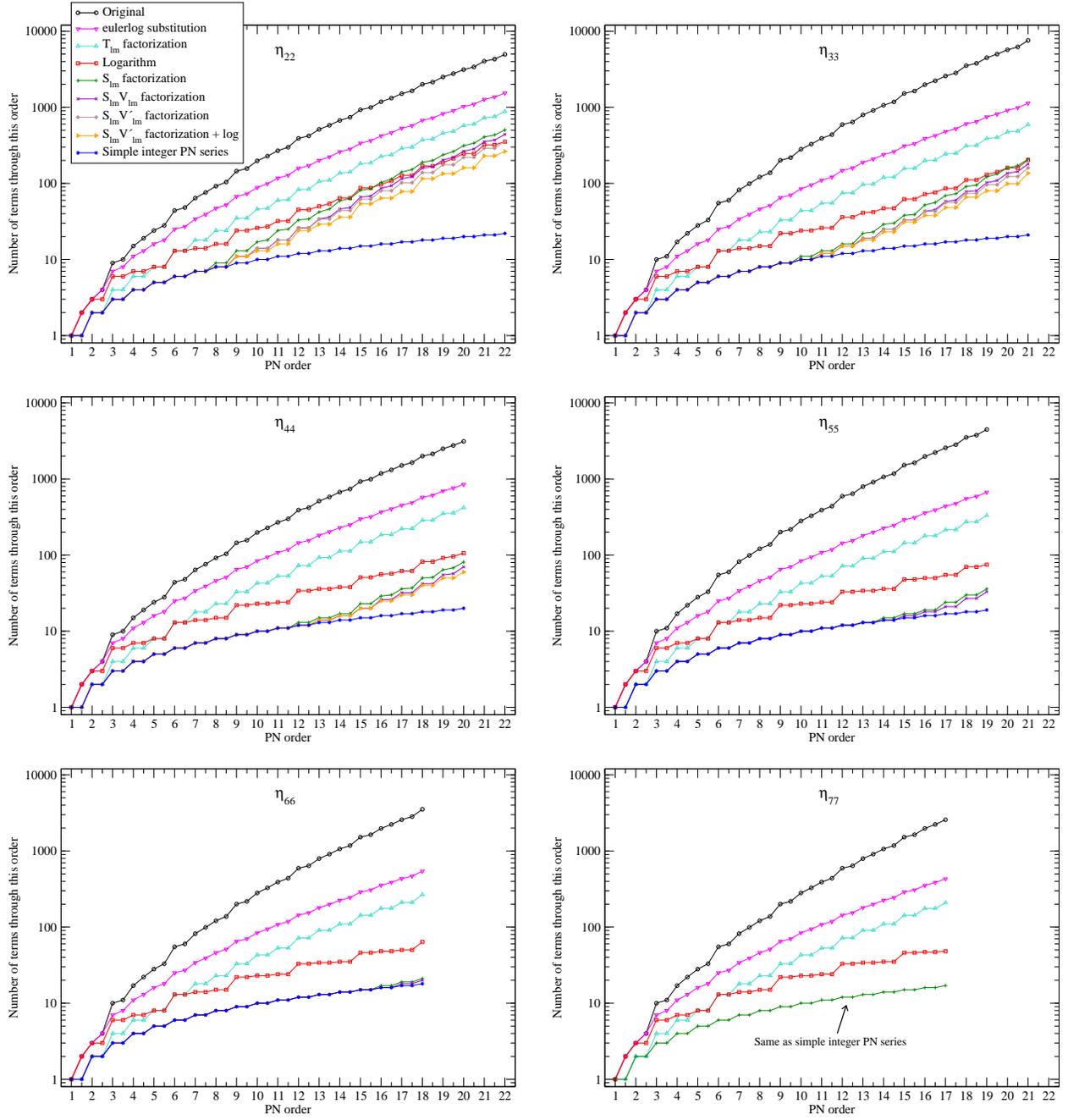

\centering
\subfloat{
\epsfig{file=eta22_lenplot.eps,width=8cm,clip=true}
}
\quad
\subfloat{
\epsfig{file=eta33_lenplot_x.eps,width=8cm,clip=true}
}\\
\subfloat{
\epsfig{file=eta44_lenplot_x.eps,width=8cm,clip=true}
}
\quad
\subfloat{
\epsfig{file=eta55_lenplot_x.eps,width=8cm,clip=true}
}\\
\subfloat{
\epsfig{file=eta66_lenplot_x.eps,width=8cm,clip=true}
}
\quad
\subfloat{
\epsfig{file=eta77_lenplot_x.eps,width=8cm,clip=true}
}
\caption{\label{fac_comp} The total number of terms up to a given PN order for the $\ell = m$ modes of the energy flux, up to $\ell = 7$, using the Fujita $22$PN energy flux expressions. (Note that we have not included the
lowest-order Newtonian term, for simplicity.) Here we show the number of terms in the expressions given by Fujita online~\cite{Fujita22PN_web} (``Original''), and then the number of terms after various simplifications or factorizations: Here ``eulerlog substitution'' denotes the result of the $\eulerlog_m(v)$ and $\log(2v^2)$ substitutions, which are also used for all the other versions shown. The other simplifications are the results of the Damour-Nagar $T_{\ell m}$ factorization,
the result of taking the logarithm (and PN expanding to the appropriate order), and the various factorizations discussed in the text, viz., $S_{\ell m}$, $S_{\ell m}V_{\ell m}$, and $S_{\ell m}V'_{\ell m}$, as well as the results of taking a logarithm (and PN expanding) after performing the $S_{\ell m}V'_{\ell m}$ factorization. We also show the length of a purely integer-order PN series with
one term per coefficient, for comparison.}
\end{figure*}

\begin{figure*}[htb]
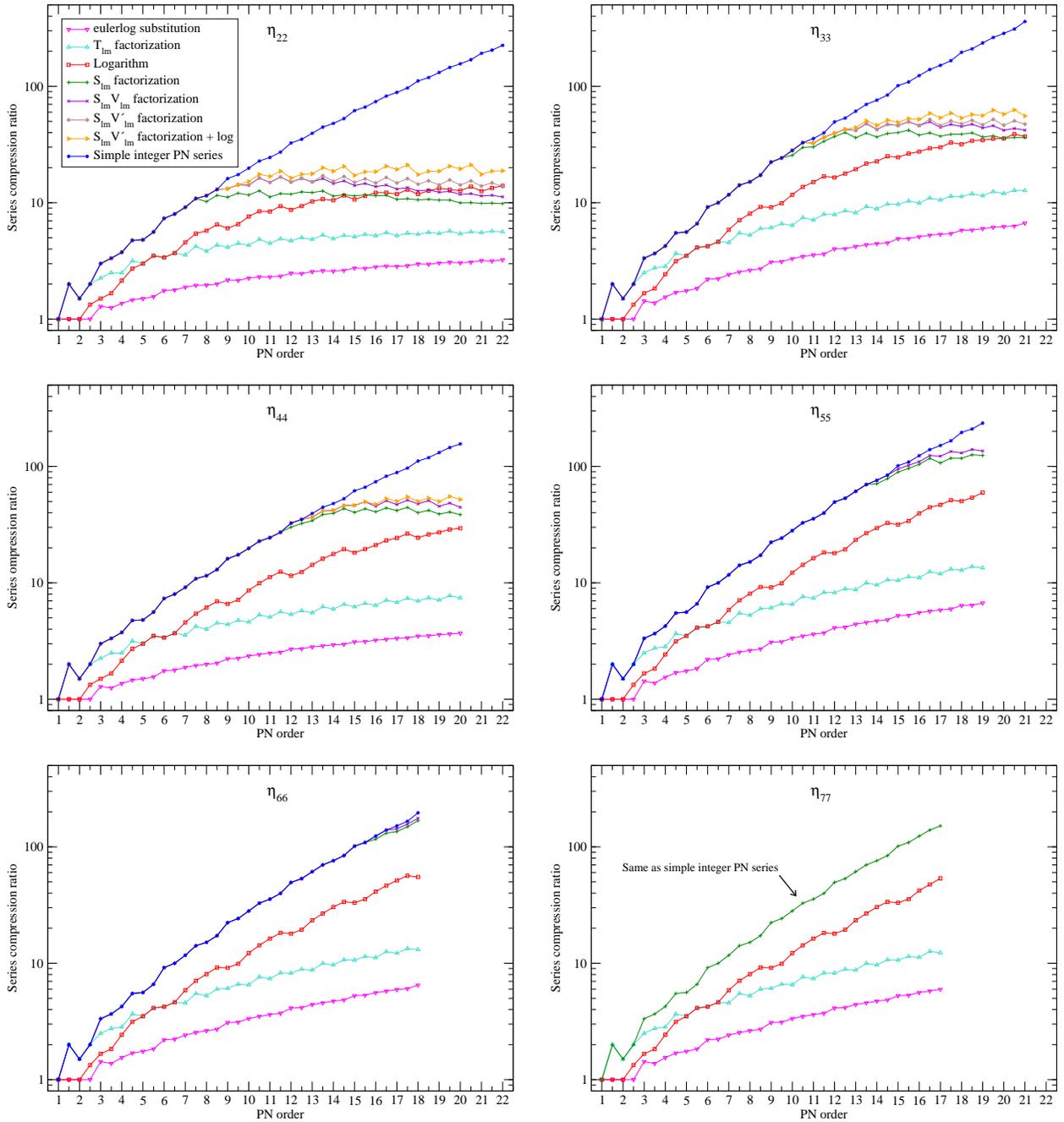

\centering
\subfloat{
\epsfig{file=eta22_compression.eps,width=8cm,clip=true}
}
\quad
\subfloat{
\epsfig{file=eta33_compression.eps,width=8cm,clip=true}
}\\
\subfloat{
\epsfig{file=eta44_compression.eps,width=8cm,clip=true}
}
\quad
\subfloat{
\epsfig{file=eta55_compression.eps,width=8cm,clip=true}
}\\
\subfloat{
\epsfig{file=eta66_compression.eps,width=8cm,clip=true}
}
\quad
\subfloat{
\epsfig{file=eta77_compression.eps,width=8cm,clip=true}
}
\caption{\label{compress_comp} The series compression ratio for various PN orders (i.e., the number of terms in the simplified series up to a given order divided by the number of terms in the original series up to that order) for the various simplifications and for a simple integer PN series, with the same notation and comments as for Fig.~\ref{fac_comp}}
\end{figure*}

We illustrate the overall performance of the various simplifications of the modes of the energy flux in Figs.~\ref{fac_comp} and~\ref{compress_comp}, where we show the total number of terms to a
given PN order as well as the ratio of this total number with that of the original series.\footnote{Note that the fractional reduction in file size when one outputs the {\sc{Mathematica}} expressions as plain text is a bit larger than the fractional reduction in the number of terms. Also note that the fractional reduction in file size one obtains using, e.g., the standard {\sc{unix}} {\sc{gzip}} compression program on the original is approximately the same as that from the eulerlog substitution, as might be expected. One gets an additional reduction in file size of a factor of $\sim 1.5$ to $\sim 3$ upon applying {\sc{gzip}} to the simplified data, with the smallest reduction for the simplest cases, as expected.} We only show results for the $\ell = m$ modes, since the simplification is the same for all
modes with the same value of $\ell$. (Of course, the modes with $\ell + m$ odd have one fewer PN order known than those with $\ell + m$ even, for a given $\ell$.) We only show the modes up to $\ell = 7$ since
the modes with $\ell > 7$ have the same simplification as those with $\ell = 7$ (except for the decrease in the number of PN coefficients known with increasing~$\ell$).

N.B.: Taking
the logarithm removes $6$ more terms total than the $S_{\ell m}V'_{\ell m}$ factorization for $\eta_{22}$ and $4$ more terms than the $S_{\ell m}$ factorization for $\eta_{33}$, though the specific complexity each simplification removes is different. We do not show the result of the pure $S_{\ell m}V'_{\ell m}$ factorization for $\eta_{44}$, to avoid a cluttered plot, since it only differs from the $S_{\ell m}V_{\ell m}$ factorization by removing two more terms at the two highest PN orders shown. Similarly, we do not show the result of expanding the logarithm of the $S_{\ell m}V_{\ell m}$ factorization for $\eta_{55}$, since it only differs from the plain $S_{\ell m}V_{\ell m}$ factorization by removing one additional term at the highest PN order shown. Additionally, we do not show the $S_{\ell m}V'_{\ell m}$ factorization for $\ell \geq 5$ and the $S_{\ell m} V_{\ell m}$ factorization for
$\ell = 7$, since the coefficients entering them cannot be fixed using the $22$PN energy flux expressions for those modes. (In particular, there are no transcendentals in the expansion of $V_{7 m}$ to $17$PN, the highest order present in the $22$PN energy flux expression for $\bar{\eta}_{77}$.) Note that both the $S_{\ell m}$ and $S_{\ell m}V_{\ell m}$ factorizations have the same number of terms as the purely integer-order PN series
up to a certain PN order, which is higher than the order known from Fujita's $22$PN calculation~\cite{Fujita22PN} for $\ell\geq 7$. Also, the jumps in the number of terms every $3$PN (particularly noticeable in the logarithm) are due to the fact that the PN expansion of $\nu$ is a power series in $v^6$, so that the complexity of the PN expansion of the modes increases significantly every $3$PN.

We now compare the results of the different simplifications more explicitly for $\eta_{22}$. While we only show explicit results for $\eta_{22}$ here, all the other modes exhibit the same general
sort of complexity: The primary difference for the modes with larger $\ell$ is that the complexity that is only incompletely removed by $V_{\ell m}$ (or even $V'_{\ell m}$) only enters at increasingly higher powers of $v$ as $\ell$ increases. We present most results to $O(v^{25})$ to illustrate what happens for the terms beyond which even the $S_{\ell m}V'_{\ell m}$ factorization does not remove all the transcendentals or the logarithms and odd powers of $v$. However, we only present the original version (with the simplifying substitutions) to $O(v^{14})$, the same order to which Fujita gives the full energy flux in~\cite{Fujita22PN}, and the $T_{\ell m}$ factorized versions to $O(v^{18})$, in order not to burden the reader with excessively lengthy expressions: Fujita's original expression for $\eta_{22}$ through
$O(v^{44})$ is available online at~\cite{Fujita22PN_web}, and we give the full expressions with the $\eulerlog_m(v)$ and $\log(2v^2)$ substitutions and the $T_{\ell m}$ factorization in the electronic material accompanying this paper~\cite{data_URL}. (Note also that the $T_{\ell m}$ factorization has the same complexity as the factorized gravitational wave amplitude $\rho_{\ell m}$, which is also given
online by Fujita.) We also do not show the $S_{\ell m}V'_{\ell m}$ factorization, since it only differs from the $S_{\ell m}V_{\ell m}$ factorization at higher orders than we display here. Similarly, to avoid unnecessary repetition, we do not explicitly show the results of combining the logarithm with the $S_{\ell m}V_{\ell m}$ factorization, since the reduction in complexity is only slightly more than the reduction one obtains from the $S_{\ell m}V_{\ell m}$ factorization on its own, to the order shown. Again, the full expressions for all these quantities are given in the electronic material accompanying this paper~\cite{data_URL}.

For the original quantity, we have 
\begin{widetext}
\<
\begin{split}
\eta_{22} &= 1 - \frac{107}{21} v^2 + 4 \pi  v^3 + \frac{4784}{1323}v^4 - \frac{428}{21}\pi  v^5 + \left[\frac{99210071}{1091475}+\frac{16}{3} \pi ^2 - \frac{1712}{105}\eulerlog_2(v)\right]v^6 +\frac{19136}{1323} \pi  v^7\\
&\quad + \left[-\frac{27956920577}{81265275} -\frac{1712 }{63} \pi ^2+ \frac{183184}{2205}\eulerlog_2(v)\right]v^8
+ \left[\frac{396840284}{1091475}-\frac{6848}{105} \eulerlog_2(v)\right]\pi v^9+\\
&\quad + \left[\frac{187037845924}{6257426175} + \frac{76544 }{3969}\pi ^2-\frac{8190208}{138915}\eulerlog_2
   (v)\right]v^{10}
   +\left[-\frac{111827682308 }{81265275} + \frac{732736}{2205}\eulerlog_2
   (v)\right]\pi v^{11} \\
   &\quad + \biggl[\frac{139638221186546204}{29253467368125} +\frac{295709968}{654885} \pi ^2 - \frac{256 }{45}\pi
   ^4 -\frac{27392 }{105}\zeta (3) - \left(\frac{36117727568}{22920975} + \frac{27392 }{315}\pi^2\right)\eulerlog_2(v)\\
   &\quad + \frac{1465472 }{11025}\eulerlog_2^2(v)\biggr]v^{12}
   + \left[\frac{748151383696 }{6257426175}-\frac{32760832}{138915}\eulerlog_2(v)\right]\pi v^{13}
   + \biggl[-\frac{19222892871566153708684}{1365639617146179375}\\
   &\quad -\frac{406031680304 }{243795825}\pi ^2+\frac{27392 }{945}\pi ^4+\frac{2930944 }{2205}\zeta (3)+\left(\frac{51936991437808}{8532853875}+\frac{2930944}{6615} \pi ^2\right) \eulerlog_2 (v)\\
   &\quad -\frac{156805504}{231525} \eulerlog_2^2(v)\biggr]v^{14} + O\left(v^{15}\right).
   \end{split}
   \?

[N.B.: The ``Original'' traces given in Fig.~\ref{fac_comp} count the terms in the original expressions given by Fujita~\cite{Fujita22PN, Fujita22PN_web}, which do not use the $\eulerlog_m(v)$ or $\log(2v^2)$
substitutions. One can obtain the original $\eta_{22}$
from the simplified version given here by writing out $\eulerlog_2(v) =  \gamma + 2\log 2 + \log v$ and $\log(2v^2) = \log 2 + 2\log v$ (though the latter does not appear to the order shown), and expanding the squares and products involving these terms, so, e.g., the $v^{12}$ and $v^{14}$ coefficients each have $16$ terms in the original version.]
We then have
\<
\begin{split}
\frac{\eta_{22}}{|T_{22}|^2} &= 1-\frac{107 }{21}v^2 +\frac{4784 }{1323}v^4+ \left[\frac{77380571}{1091475} -\frac{1712}{105} \eulerlog_2(v)\right]v^6+ \left[-\frac{19675602077}{81265275} + \frac{183184 }{2205}\eulerlog_2(v)\right]v^8\\
&\quad+ \left[-\frac{265502242076}{6257426175}-\frac{8190208}{138915} \eulerlog_2 (v)\right]v^{10}+\biggl[\frac{96287266208715704}{29253467368125} -\frac{366368 }{11025}\pi ^2-\frac{27392 }{105}\zeta (3)\\
&\quad -\frac{28643306768 }{22920975}\eulerlog_2 (v)+ \frac{1465472 }{11025}\eulerlog_2 ^2(v)\biggr]v^{12} +
   \biggl[-\frac{12164707404850205644184}{1365639617146179375} +\frac{39201376 }{231525}\pi ^2\\
   &\quad +\frac{2930944 }{2205}\zeta(3)+\frac{37759374165808 }{8532853875}\eulerlog_2 (v) -\frac{156805504 }{231525}\eulerlog_2 ^2(v)\biggr]v^{14}+
    \biggl[-\frac{47549782802370518469284}{9559477320023255625}\\
    &\quad -\frac{1752704512 }{14586075}\pi^2 -\frac{131043328}{138915} \zeta (3) +\frac{46374364943936}{131405949675}\eulerlog_2(v) + \frac{7010818048}{14586075}\eulerlog_2^2(v) -\frac{512}{15} \log (2 v^2)\biggr]v^{16}\\
    &\quad + \biggl[\frac{373743318483721726610648630768}{3083648396506501683234375}-\frac{32946911447392 }{12033511875}\pi^2 -\frac{5861888 }{165375}\pi ^4 -\frac{2581241112832}{114604875} \zeta (3) +\frac{438272}{105} \zeta (5)\\
    &\quad +\left(-\frac{189488492579157761216}{3071614073653125}+\frac{627222016}{1157625} \pi ^2 + \frac{46895104 }{11025}\zeta (3)\right)\eulerlog_2(v)  +\frac{131787645789568 }{12033511875}\eulerlog_2^2(v)\\
    &\quad -\frac{2508888064 }{3472875}\eulerlog_2^3(v)+\frac{11264}{315} \log (2 v^2)\biggr]v^{18} + O\left(v^{20}\right),
\end{split}
\?

(N.B.: The $T_{\ell m}$ factorization only removes the odd powers of $v$ through $v^{7+6\ell}$, so the first odd power of $v$ in $\eta_{22}/|T_{22}|^2$ is $v^{21}$.)
\<
\begin{split}
\frac{\eta_{22}}{|S_{22}|^2} &= 1-\frac{107 }{21}v^2+\frac{4784 }{1323}v^4+\frac{99210071 }{1091475}v^6-\frac{27956920577 }{81265275}v^8+\frac{187037845924 }{6257426175}v^{10}+\frac{139638221186546204
   }{29253467368125}v^{12}\\
   &\quad -\frac{19222892871566153708684 }{1365639617146179375}v^{14}+ \left[-\frac{53449637268712260375284}{9559477320023255625}-\frac{512}{15}\log
   (2v^2)\right]v^{16}\\
   &\quad +\left[\frac{590730250424481655118765186768}{3083648396506501683234375}-\frac{114688  }{1605}\eulerlog_2(v) +\frac{11264}{315}\log (2v^2)\right]v^{18}\\
   &\quad +
   \left[-\frac{3235369286024903603645361349174816}{6854950385433953241830015625}+\frac{16384  }{45}\eulerlog_2(v) -\frac{42752}{405}\log (2v^2)\right]v^{20}-\frac{16384}{75} \pi 
   v^{21}\\
   &\quad + \biggl[-\frac{8377976958392263106467167178591936}{22392837925750913923311384375}-\frac{876544 }{4725}\pi ^2-\frac{16384}{15} \zeta
   (3)-\frac{78381056  }{303345}\eulerlog_2(v)  - \frac{146048512}{40425}\log (2v^2)\\
   &\quad+\frac{438272 }{1575}\log
   ^2(2v^2)\biggr]v^{22}+\frac{1753088 }{1575}\pi  v^{23}+ \biggl[\frac{431571188712518783017237317313849608070192}{64125832996859366542516201416796875} -\frac{8667136 }{19845}\pi ^2\\
   &\quad +\frac{32768}{33705}\zeta (3)-\frac{362080557826048 }{26777779875}\eulerlog_2(v) + \frac{131072 }{225}\eulerlog_2^2(v)+
   \frac{262144}{225} \eulerlog_2(v)\log (2 v^2) \\
   &\quad+\frac{3218927074816}{1489863375}
   \log (2 v^2) -\frac{9641984 }{33075}\log ^2(2 v^2)\biggr]v^{24}-\frac{78381056 }{99225}\pi 
   v^{25}+O\left(v^{26}\right),
\end{split}
\?
\<
\label{etaSV22}
\begin{split}
\frac{\eta_{22}}{|S_{22}V_{22}|^2} &= 1-\frac{107 }{21}v^2+\frac{4784 }{1323}v^4+\frac{99210071 }{1091475}v^6-\frac{27956920577 }{81265275}v^8+\frac{18611386050668 }{669544600725}v^{10}+\frac{139950258171806204
   }{29253467368125}v^{12}\\
   &\quad -\frac{2057955690896253972269188 }{146123439034641193125}v^{14}-\frac{5913747619091360176272988 }{1022864073242488351875}v^{16}\\
   &\quad + \left[\frac{63445764566704501684220321444176}{329950378426195680106078125} -\frac{114688  }{1605}\eulerlog_2(v) -\frac{8704}{63}\log (2v^2)\right]v^{18}\\
   &\quad + \left[-\frac{37046395989475217401031049669895828384}{78482326962833330665711848890625} +\frac{16384 }{45} \eulerlog_2(v) +\frac{70912}{3969}\log (2v^2)\right]v^{20}\\
   &\quad +  \left[-\frac{919480615890647231616774790102430912}{2396033658055347789794318128125}-\frac{876544}{4725}  \pi^2 -\frac{78381056}{303345} \eulerlog_2(v)-\frac{110035171328}{350363475}\log(2v^2)\right]v^{22}\\
   &\quad + \biggl[\frac{4970118699958639953829221250019069029006428208}{7341766619810428875
   45267990020907421875} + \frac{43368448}{99225}\pi^2 -\frac{187547648}{33705}\zeta (3)\\
   &\quad -\frac{81953900429312}{5355555975}\eulerlog_2(v) + \frac{131072}{225} \eulerlog_2^2(v)+\frac{262144}{225}\eulerlog_2(v) \log(2v^2)-\frac{165767673909248 }{13408770375}\log(2v^2)\\
   &\quad +\frac{7450624
  }{6615} \log^2(2v^2)\biggr]v^{24}+O\left(v^{26}\right),
   \end{split}
\?
\<
\label{leta22}
\begin{split}
\log\eta_{22} &= -\frac{107 }{21}v^2+4 \pi  v^3-\frac{24779}{2646}v^4+\left[\frac{166117624}{2546775} -\frac{8 }{3} \pi ^2 -\frac{1712}{105}\eulerlog_2(v)\right]v^6 +\frac{949963732033 }{25029704700}v^8+\frac{394091579209 }{11945995425}v^{10}\\
&\quad + \left[\frac{93938578959284551438}{116107011984088125} -\frac{366368 }{11025}\pi ^2 +\frac{64 }{45}\pi^4 -\frac{27392}{105} \zeta (3) -\frac{108494912 }{1157625}\eulerlog_2(v)\right]v^{12}\\
   &\quad +\frac{47297778686376177755969
  }{82661362708436386875} v^{14} + \left[\frac{18189001464420142574516153}{20235501591025227507000}-\frac{512 }{15}\log (2v^2)\right]v^{16}\\
  &\quad + \biggl[\frac{3554229825508450422696607122848}{244082630769630017849859375} -\frac{46435822336}{121550625} \pi^2-\frac{5861888 }{165375}\pi ^4-\frac{4096 }{2835}\pi ^6-\frac{111910912 }{27783}\zeta (3)+\frac{438272 }{105}\zeta (5)\\
  &\quad -\frac{193586530718464 }{140390971875}\eulerlog_2(v) -\frac{8704}{63}\log (2v^2)\biggr]v^{18} +
   \biggl[\frac{11032318346990181267793875894596613233}{888771737273053773616228845843750}\\
  &\quad-\frac{2723072}{3969} \log (2v^2)\biggr]v^{20}-\frac{16384  }{75} \pi v^{21} +
   \biggl[\frac{211320713200843785527303127211486164929}{9332103241367064622970402881359375} -\frac{876544 }{4725}\pi^2-\frac{16384 }{15}\zeta (3)\\
   &\quad-\frac{401855693056}{114604875} \log (2v^2) + \frac{438272}{1575} \log ^2(2v^2)\biggr]v^{22} + \biggl[\frac{1736732120102373166012533790428894237181748497283}{5499823952034433050868796175173372213671875}\\
   &\quad -\frac{2558186604995072}{327578934375} \pi^2 -\frac{471366904832 }{607753125}\pi^4 +\frac{187580416}{2083725}\pi^6 + \frac{8192 }{4725}\pi^8 -\frac{9839419963916288 }{140390971875}\zeta (3)\\
   &\quad  -\frac{12367511552 }{1157625}\zeta (5) -\frac{7012352}{105} \zeta (7) -\frac{4136838568957390627328}{132802137890296875} \eulerlog_2(v) + \frac{131072 }{225}\eulerlog_2^2(v)\\
   &\quad +\frac{262144}{225}\eulerlog_2(v)\log (2 v^2)-\frac{1164753374080768}{93861392625} \log (2 v^2)+\frac{7450624 }{6615}\log^2(2v^2)\biggr]v^{24}+O\left(v^{26}\right).
\end{split}
\?
\end{widetext}
If one combines the logarithm and the $S_{\ell m}V_{\ell m}$ factorization [i.e., considers $\log(\eta_{22}/|S_{22}V_{22}|^2)$], then the only difference in complexity through the order shown here [$O(v^{24})$]
is the removal of the $\eulerlog_2(v)$ terms in the coefficients of $v^{20}$ and $v^{22}$. (Since $24$ is divisible by $6$, the logarithm produces no extra simplification of the $v^{24}$ term.)

It is clear from Figs.~\ref{fac_comp} and~\ref{compress_comp} that there is some structure in the PN expansion of the small-$\ell$ modes of the energy flux, particularly for $\ell = 2$, that is simplified by taking the logarithm but is not captured in $V_{\ell m}$, or even $V'_{\ell m}$, due to the more
rapid increase in the number of terms at higher orders for the $S_{\ell m}V'_{\ell m}$ factorization compared with the result of taking the logarithm. In particular, comparing $\log\eta_{22}$ [Eq.~\eqref{leta22}] with $\eta_{22}/|S_{22}V_{22}|^2$ [Eq.~\eqref{etaSV22}], one sees that the $S_{\ell m}V_{\ell m}$ factorization removes all the transcendentals from $\eta_{22}$ at lower PN orders, while taking the logarithm
only confines the transcendentals to certain terms (with the highest complexity in powers of $v$ that are divisible by $6$), though it removes the $\eulerlog_2(v)$ terms in the $v^{20}$ and $v^{22}$ coefficients. This removal of $\eulerlog_m(v)$ contributions in coefficients of powers that are not divisible by $6$ in the expansion of the logarithm continues at higher orders and for other modes, as well. Additionally, the expansion of the logarithm also only has odd powers of $v$ of the form $v^{9+6(\ell + n)}$, $n\in\N_0$. 

While the $S_{\ell m}V'_{\ell m}$ factorization successfully removes \emph{all} the odd powers of $v$, it does not accomplish the complete removal of the $\eulerlog_m(v)$ terms in the coefficients of powers of $v$ that are not divisible by $6$ that is achieved by expanding the logarithm. This is likely another indicator of further structure in the expansion that is not being captured by $S_{\ell m}V'_{\ell m}$: For instance, the coefficients of $\eulerlog_2^2(v)v^{24}$ and $\eulerlog_2(v)\log(2v^2)v^{24}$ in $\eta_{22}/|S_{22}V_{22}|^2$ are $8/5$ and $16/5$ times the coefficient of $\eulerlog_2(v)v^{20}$, respectively, and this coefficient itself has the simple form $2^{14}/3^25^1$. Moreover, the $\pi v^{21}$ and $\zeta(3)v^{22}$ terms in $\eta_{22}/|S_{22}|^2$, which \emph{are} removed by the $V_{\ell m}$ factorization, also have coefficients that are very closely related to the coefficient of  $\eulerlog_2(v)v^{20}$, viz., $-3/5$ and $-3$ times it, respectively. Also note that the $\eulerlog_m(v)$ contributions that are not removed by the $S_{\ell m}V'_{\ell m}$ factorization first arise at one PN order higher for each increase in $\ell$ by $1$. In particular, this means that the $22$PN energy flux expression for $\bar{\eta}_{6m}/|S_{6m}V_{6m}|^2$ only differ from a pure integer PN series with rational coefficients by one or two (depending on whether $m$ is odd or even, respectively) $\log(2v^2)$ terms in the final one or two PN orders, which could be removed with the $q_{\ell m}$ substitution [Eq.~\eqref{qlm_subs}]. 

Finally, let us consider the overall simplification of the three highest-order coefficients in the $22$PN expansion of $\eta_{22}$ produced by the $S_{\ell m}V'_{\ell m}$ factorization, expanding the logarithm, and combining the two: The original
version of $\eta_{22}$ has $171$ terms each in the coefficients of $v^{42}$ and $v^{44}$ and $93$ terms in the coefficient of $v^{43}$, even after the $\eulerlog_m(v)$ and $\log(2v^2)$ substitutions. The $S_{\ell m}V'_{\ell m}$ factorization and logarithm both remove the $v^{43}$ coefficient completely (though recall that the $S_{\ell m}V'_{\ell m}$ factorization removes \emph{all} the odd powers of $v$, while the logarithm only removes those that are not of the form $v^{9+6(\ell + n)}$, $n\in\N_0$, so, e.g., the $v^{39}$ term remains). However, they simplify the $v^{42}$ and $v^{44}$ coefficients slightly differently: The $S_{\ell m}V'_{\ell m}$ factorization gives $70$ terms in both coefficients, while the logarithm gives $75$ terms in the coefficient of $v^{42}$ and only $33$ in the coefficient of $v^{44}$. Combining the two simplifications gives (as might be expected) the minimum number of terms in both coefficients, viz., $70$ for $v^{42}$ and $33$ for $v^{44}$.

Since $42$ is divisible by $6$, the logarithm does not reduce the complexity of zeta values present in the coefficient of $v^{42}$ in $\eta_{22}$: $\zeta(n)$ is present through $n = 14$ in the logarithm, as it is in the original, while the maximum $n$ for which $\zeta(n)$ is present with the $S_{\ell m}V'_{\ell m}$ factorization is $9$. However, with both the $S_{\ell m}V'_{\ell m}$ factorization and the logarithm, the maximum power of $\eulerlog_2(v)$ in the coefficient of $v^{42}$ decreases from $7$ to $5$, as does the maximum order of the products present, though neither of these simplifications reduce the maximum power of $\log(2v^2)$ present from $5$. For the $v^{44}$ coefficient of $\eta_{22}$, the $S_{\ell m}V'_{\ell m}$ factorization gives the same simplification as for $v^{42}$, while the logarithm gives a much greater simplification, removing \emph{all} the powers of $\eulerlog_2(v)$, in addition to giving the same simplification of the values of $\zeta(n)$ present as the $S_{\ell m}V'_{\ell m}$ factorization. Of course, for modes with higher
$\ell$, for which the $-\nu-1$ terms in the MST formalism make less of a contribution, the simplification of the highest terms produced by the $S_{\ell m}V'_{\ell m}$ factorization is much more considerable, as is
illustrated in Figs.~\ref{fac_comp} and~\ref{compress_comp}.

%%%%%%%%%%%%%%%%%%%%%%%%%
\section{Convergence of the simplified modes of the energy flux}\label{sec:comp}
%%%%%%%%%%%%%%%%%%%%%%%%%

We now wish to see how these various factorizations and resummations affect the convergence of the series, so we consider a few illustrative cases, saving a more detailed investigation and comparison with other
resummation methods (e.g., Pad{\'e}~\cite{DIN} and Chebyschev~\cite{CP}) to future work. Specifically, we compare the convergence of $\eta_{22}$, $\eta_{21}$, $\eta_{33}$, and $\eta_{55}$ for orbits at two relatively small radii, viz., $r_0 = 6M$, the Schwarzschild innermost stable circular orbit (ISCO), and $10M$, providing a rather stringent test of the convergence. In both cases, we compare with the fluxes calculated numerically by Fujita and Tagoshi~\cite{FT2004}: They give the modes of the flux to $11$ digits through $\ell = 6$ at the ISCO in their Table~III, and the modes of the flux to $17$ digits through $\ell = 7$ for $r_0 = 10M$ in their Table~VIII, where they
compare with the data calculated by Tagoshi and Nakamura by a different method~\cite{TN} and find that these expressions are valid to $13$ or $14$ digits.
\begin{figure*}[htb]
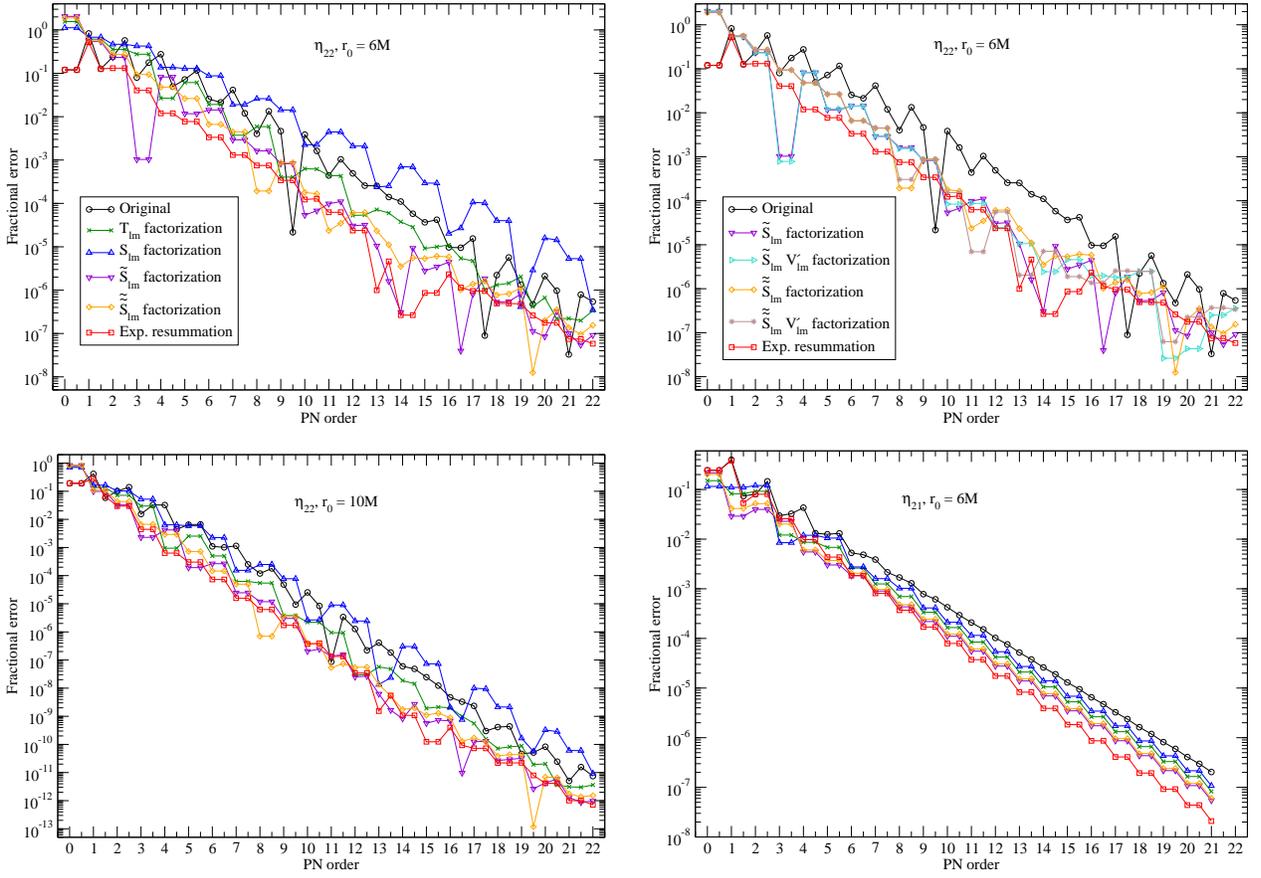

\centering
\subfloat{
\epsfig{file=eta22_conv_comp_r6_corr.eps,width=8cm,clip=true}
}
\quad
\subfloat{
\epsfig{file=eta22_conv_comp2_r6_corr.eps,width=8cm,clip=true}
}\\
\subfloat{
\epsfig{file=eta22_conv_comp_r10FT_corr.eps,width=8cm,clip=true}
}
\quad
\subfloat{
\epsfig{file=eta21_conv_comp_r6_corr.eps,width=8cm,clip=true}
}\\
\caption{\label{conv_comp1} Convergence of different factorizations and resummations for $\eta_{22}$ and $\eta_{21}$, computed for orbital radii of $r_0 = 6M$ (the ISCO) for both modes and also for $r_0 = 10M$ for $\eta_{22}$ (the convergence of the various versions for $\eta_{21}$ for this radius are the same as at the ISCO, except more rapid). We compare to the fluxes computed numerically by Fujita and Tagoshi~\cite{FT2004}. The notation for the different factorizations is that introduced in the text. The legend in the upper left-hand figure also applies to the bottom two
figures; the legend in the upper right-hand figure just applies to that figure, which shows the effects of adding in the $V'_{\ell m}$ factorization. Note also that the horizontal scales of the plots are all the same, but the vertical scales differ.}
\end{figure*}

\begin{figure*}[htb]
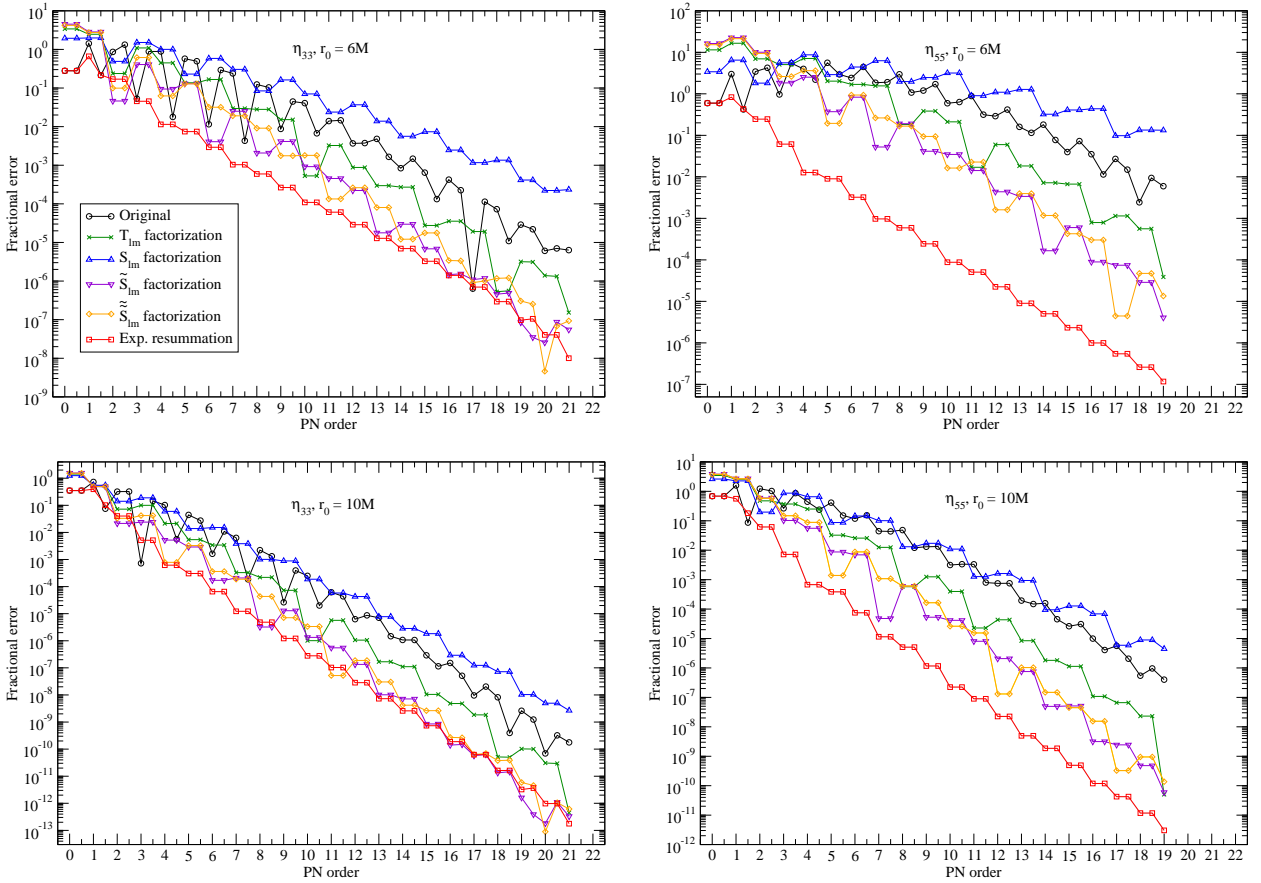

\centering
\subfloat{
\epsfig{file=eta33_conv_comp_r6_corr.eps,width=8cm,clip=true}
}
\quad
\subfloat{
\epsfig{file=eta55_conv_comp_r6_corr.eps,width=8cm,clip=true}
}\\
\subfloat{
\epsfig{file=eta33_conv_comp_r10FT_corr.eps,width=8cm,clip=true}
}
\quad
\subfloat{
\epsfig{file=eta55_conv_comp_r10FT_corr.eps,width=8cm,clip=true}
}\\
\caption{\label{conv_comp2} Convergence of different factorizations and resummations for $\eta_{33}$ and $\eta_{55}$, computed for orbital radii of $r_0 = 6M$ (the ISCO) and $r_0 = 10M$.
The notation and other comments are the same as for Fig.~\ref{conv_comp1}, except that we do not show any results for the factorizations involving $V'_{\ell m}$ here, for simplicity---including $V'_{\ell m}$ only changes the results at the highest orders, where the difference in behavior is of the same sort
seen for $\eta_{22}$, $r_0 = 6M$ in Fig.~\ref{conv_comp1}.}
\end{figure*}

Since we find that the $S_{\ell m}$ factorization does not improve convergence (and indeed makes it less rapid than that of the original series in many cases), while the $T_{\ell m}$ factorization does improve convergence (though not as much as the exponential resummation), we also consider two alternatives to $S_{\ell m}$ that include the $\ell$ terms accompanying $\nu$ in the gamma functions (since similar factors are present in $T_{\ell m}$), viz.,
\begin{subequations}
\label{St}
\begin{align}
\tilde{S}_{\ell m} &:= (2m v)^{\bar{\nu}_{\ell m}(v)}e^{\pi m v^3}\frac{(2\ell + 1)!}{(\ell - 2)!}\nonumber\\
&\quad\;\times \frac{\Gamma[-1 + \ell + \bar{\nu}_{\ell m}(v) - 2\ri m v^3]}{\Gamma[2+2\ell + 2\bar{\nu}_{\ell m}(v)]},\\
\tilde{\tilde{S}}_{\ell m} &:= (2m v)^{\bar{\nu}_{\ell m}(v)}e^{\pi m v^3}\frac{(2\ell)!}{\ell!}\frac{\Gamma[1 + \ell + \bar{\nu}_{\ell m}(v) - 2\ri m v^3]}{\Gamma[1+2\ell + 2\bar{\nu}_{\ell m}(v)]}.
\end{align}
\end{subequations}
Specifically, $\tilde{S}_{\ell m}$ contains the constants and $\ell$ terms in the arguments of the gamma functions that come from considering the gamma functions present in $|R_\text{C}^\nu/A_+^\nu|$, which is the portion of the MST formalism that leads to $S_{\ell m}$, as discussed in Sec.~\ref{fac_disc}. We also add an overall scaling (the factorials) to make the first term in the series unity. For $\tilde{\tilde{S}}_{\ell m}$, we just substitute $\bar{\nu}_{\ell m}(v) \to \nu = \ell + \bar{\nu}_{\ell m}(v)$ inside the gamma functions in $S_{\ell m}$ and make the same sort of overall scaling. (These overall scalings do not affect the numerical convergence, though they make the coefficients in the expansion simpler and thus slightly speed up calculations of the factorization.)

We illustrate the convergence of the series with the various factorizations and resummations in Figs.~\ref{conv_comp1} and~\ref{conv_comp2}. For the various factorizations, we compute the value of the flux by multiplying the value of the full factorization with no expansion with the value of the expansion of the factorized flux to the given order. Of course, we must also calculate the value of $\nu$ used in the factorizations
we introduce, but here we merely use the
$O(\epsilon^{18})$ [i.e., $O(v^{54})$] expansions provided to us by Abhay G.\ Shah: The convergence of the PN expansion of $\nu$ is rapid and monotonic, so the fractional error in the value for $\nu$ we obtain from it (measured from self-convergence) is
always at least two orders of magnitude below the minimum error we find for a given mode of the flux. Additionally, we can compare with the values of $\nu$ given to $14$ digits in Table~II of Fujita and Tagoshi~\cite{FT2005} for $\ell = 2$.
Here we find that the $O(\epsilon^{18})$ expansion reproduces all $14$
digits (up to the final digit, where the discrepancy may be due to rounding) for $M\omega = 0.1$ (corresponding to a circular orbit at a radius of $r_0 \simeq 7.4M$ for $m = 2$; recall that $\omega = m\Omega$). The $O(\epsilon^{18})$ expansion does not reproduce all $14$ digits for the other values of $M\omega$ in that table, but all of these correspond to orbits inside the ISCO.

We find that in all cases we consider that the exponential resummation produces the cleanest and fastest convergence, in some cases improving the convergence by more than four orders of magnitude (for modes with large $m$),
compared with the original series. The $\tilde{S}_{\ell m}$ and $\tilde{\tilde{S}}_{\ell m}$ factorizations also improve the convergence, in some cases almost as much as the exponential resummation, though they do so far less cleanly. While the specifics of the convergence of these two factorizations differ substantially, neither is clearly preferable. Adding in the $V'_{\ell m}$ factorization modifies the specifics of the convergence, but again produces no clear improvement. (We only show this for $\eta_{22}$ at $r_0 = 6M$, but the effects are similar for the other cases, though the differences are less pronounced overall.) If one combines the exponential resummation with the $\tilde{S}_{\ell m}$ or $\tilde{\tilde{S}}_{\ell m}$ factorizations, one obtains much the same results as the exponential resummation on its own, so we do not show this. The $T_{\ell m}$ factorization also improves the convergence, though less than the $\tilde{S}_{\ell m}$ and $\tilde{\tilde{S}}_{\ell m}$ factorizations. Finally, the $S_{\ell m}$ factorization worsens the series' convergence in all cases we show, except for $\eta_{21}$.

It is not clear why the different modes exhibit the significant differences in convergence seen in Figs.~\ref{conv_comp1} and~\ref{conv_comp2}. In particular, the large dip in the convergence of $\log\eta_{22}$ for $r_0 = 6M$ does not occur at the PN order where one first gets contributions from the $-\nu - 1$ terms in the MST formalism, which is $8$PN, at which point one might expect something unusual might happen, but rather somewhat later, between about $12$PN and $16$PN. While $12$PN is indeed the point at which $\log\eta_{22}$ starts to display large increases in complexity [see Eq.~\eqref{leta22}], the same is true for $\log\eta_{21}$, and we do not see similar
behavior there. Indeed, it is surprising that the convergence of $\eta_{21}$ is so monotonic, compared with $\eta_{22}$. Some of this may be due to the fact that the expansion parameter for many of the quantities in the MST formalism is $\epsilon = 2mv^3$, which becomes smaller as $m$ decreases (for a given $v$), but this is surely not all. However, this argument with $\epsilon$ may help explain why the original series for the modes with higher $m$ converge significantly less rapidly than those with small $m$;  cf., e.g., $\eta_{55}$ with the other modes shown in Figs.~\ref{conv_comp1} and~\ref{conv_comp2}. For comparison, though we do not show this, the original $\eta_{51}$ converges at very close to the same rate as the exponential resummation of $\eta_{55}$ at higher orders (with a fractional error at the highest order known of $\sim 10^{-7}$ for $r_0 = 6M$), and exponential resummation makes only a very small improvement in the convergence of $\eta_{51}$. Nevertheless, the argument with $\epsilon$ does not seem to explain why the exponential resummation, in particular, is so effective in increasing the high-$m$ modes' rate of convergence.

Finally, it is also unclear why the half-integer PN contributions in the factorizations and resummations of $\eta_{21}$ have such a minor effect on the numerical value at higher orders (their effect being almost invisible on the plot)---at least some half-integer PN terms (i.e., odd powers of $v$) are present at high orders in all the versions except the factorizations involving $V'_{\ell m}$, which we do not show for $\eta_{21}$. This behavior is also seen in some of the factorizations and resummations of $\eta_{33}$, though note that the exponential resummation removes all the half-integer PN terms through $21$PN (the highest order known) except those at $1.5$, $13.5$, $16.5$, and $19.5$PN. In $\eta_{55}$, this behavior is expected, since all the factorizations and exponential resummation give a purely integer-order PN series through $19$PN (the highest order known).

%%%%%%%%%%%%%%%%%%%%%%%%%
\section{Summary and Outlook}\label{sec:concl}
%%%%%%%%%%%%%%%%%%%%%%%%%

We have presented a simplifying factorization of the spherical harmonic modes of the gravitational wave energy flux (and thus also the amplitude of the gravitational waves) emitted by a point particle in a circular orbit around a Schwarzschild black hole. 
Specifically, the simplest version of this factorization, using $S_{\ell m}$ [Eq.~\eqref{Slm}], removes all the transcendentals as well as the logarithms and odd powers of $v$ from the PN expansion of the scaled energy flux mode $\bar{\eta}_{\ell m}$ through $O(v^{7+4\ell})$, and substantially simplifies the higher terms. This means that the $S_{\ell m}$ factorization turns the $22$PN energy flux results for the modes with $\ell\geq 7$ into pure integer PN series with rational coefficients. The lack of a complete simplification of the higher orders for smaller $\ell$ can be understood in the Mano, Suzuki, and Takasugi (MST) formalism~\cite{MST1,MST2} as arising from the contributions of the $-\nu-1$ terms, which only enter at relatively high orders (and the order at which they enter increases with $\ell$)---see the discussion in Sec.~\ref{fac_disc}.

One can remove further transcendentals as well as more logarithms and odd powers of $v$ from the higher terms by making an additional factorization using $V_{\ell m}$ [Eqs.~\eqref{Vlm}], though here the factorization is more complicated, and the simplification is much less dramatic: The factorization only removes some of the terms from the $v^{8+4\ell + 6n}$ coefficients and from the remaining odd powers of $v$ in $\bar{\eta}_{\ell m}$ (though it does completely remove the first three remaining odd powers of $v$, starting with $v^{9 + 6\ell}$). Nevertheless, one can remove all the remaining odd powers of $v$ (only present in the $22$PN energy flux results for $\ell \leq 4$) by replacing a coefficient in $V_{\ell m}$ with the appropriate power series (in $v^6$), obtaining $V'_{\ell m}$, as discussed in Sec.~\ref{ssec:Vplm}. It is also possible to remove further terms from other coefficients by making another such substitution in $V_{\ell m}$, but this removes far fewer terms for each added term, so we did not pursue it extensively. However, this substitution would allow the $S_{\ell m}V'_{\ell m}$ factorization to turn the $22$PN energy flux results for the $\ell = 6$ modes into simple integer PN series with rational coefficients. Additionally, in all cases the expressions are simplified by introducing the $\eulerlog_m(v)$ function introduced by Damour, Iyer, and Nagar~\cite{DIN} and then writing the remaining $\log v$ terms in terms of $\log(2v^2)$.

We have also compared the performance of the factorizations we introduce with the Damour-Nagar tail resummation~\cite{DN2007} and the exponential resummation introduced by Isoyama~\emph{et al.}~\cite{Isoyamaetal}, both in terms of simplification of the series' analytic form as well as the improvement of its numerical convergence. Here we find that the factorizations we introduce lead to significantly
greater simplification than any of the other simplifications considered for $\ell \geq 5$, reducing the number of terms in the $22$PN energy flux expressions by up to a factor of $\sim 150$. The exponential resummation produces the next greatest simplification, followed by
the $T_{\ell m}$ factorization and then the eulerlog substitution (though note that we also employ this substitution in the other simplifications). For modes with small $\ell$, the factorizations we introduce only
reduce the number of terms by a smaller factor ($\sim 10$ for $\ell  = 2$), so the exponential resummation produces a similar (or even slightly greater) simplification than our factorizations, though the hierarchy of other simplifications remains the same. Of course, for those modes one can also combine one of our factorizations with the exponential resummation to produce a further simplification.

The exponential resummation always produces the greatest improvement in the series' numerical convergence (both in speed and monotonicity), improving the accuracy of the $22$PN energy flux expressions by more than four orders of magnitude for modes with large $m$. The factorizations we introduce also improve the speed of convergence when one includes the appropriate terms involving $\ell$ in $S_{\ell m}$ [obtaining $\tilde{S}_{\ell m}$ or $\tilde{\tilde{S}}_{\ell m}$, given in Eqs.~\eqref{St}]. In some cases this improvement is almost as much as that from the exponential resummation (though not for large $m$), though these factorizations do not improve monotonicity. The original $S_{\ell m}$ factorization (without the additional $\ell$ terms) actually reduces the speed of convergence in most cases. The
$T_{\ell m}$ factorization also improves the speed of convergence, though not as much as the factorizations we introduce, and also does not improve monotonicity.

We give expressions for these various simplifications, as well as various ancillary quantities (e.g., expressions for the factorizations and the PN expansions of $\nu$ calculated for us by Abhay G.\ Shah) in a
{\sc{Mathematica}} notebook in the accompanying electronic material~\cite{data_URL}. Specifically, for the modes with $\ell \leq 7$, we give $22$PN energy flux expressions for $\bar{\eta}_{\ell m}$ with the $\eulerlog_m(v)$ and $\log(2v^2)$ substitutions and the $T_{\ell m}$, $S_{\ell m}$, $\tilde{S}_{\ell m}$, $\tilde{\tilde{S}}_{\ell m}$, $S_{\ell m}V_{\ell m}$, $S_{\ell m}V'_{\ell m}$, $\tilde{S}_{\ell m}V'_{\ell m}$, and $\tilde{\tilde{S}}_{\ell m}V'_{\ell m}$ factorizations, in addition to the PN expansions of $\log\bar{\eta}_{\ell m}$
and $\log(\bar{\eta}_{\ell m}/|S_{\ell m}V'_{\ell m}|^2)$. We also give example code to allow users to calculate these simplifications for any of the modes provided by Fujita~\cite{Fujita22PN,Fujita22PN_web}.

These sorts of factorizations might make calculating even higher-order terms in the expansion of the energy flux more feasible, since the final result could be expressed in a considerably simpler form: Here one might also split up each mode's contribution into a sum of the $\nu$ and $-\nu-1$ pieces in $R^\text{in}_{\ell m\omega}$ [see Eq.~\eqref{Rin}], since each of these pieces is likely to factorize more easily on its own
(and could likely also obtain a better improvement of simplification and convergence from the exponential resummation). It is also possible that these sorts of
studies could yield more insight into the physical nature of the different components of the MST formalism for solving the Teukolsky equation, which, as remarked upon by Sasaki and Tagoshi~\cite{ST}, has remained relatively opaque, despite the formalism's prowess as a calculational tool. Indeed, the factorization depends crucially on the renormalized angular momentum $\nu$ introduced in the MST formalism, which, as discussed in Sec.~\ref{fac_disc}, appears to be related to tail effects.

Additionally, these sorts of studies of the mathematical structure of high-order perturbation
expansions are common in quantum field theory, where one finds deep connections to other mathematical structures: See, e.g.,~\cite{Broadhurst, Schnetz, BS}. It would be interesting to see if similar structures exist in the post-Newtonian
expansion, particularly since it can be written in terms of the same sorts of loop integrals studied in quantum field theory, as noted by Bini and Damour~\cite{BD3}. It would also be interesting to relate the factorizations found here (and the MST formalism in general) to the Heun function solutions to the Regge-Wheeler and Teukolsky equations found by Fiziev~\cite{Fiziev2006,Fiziev2010}.

There are also many more potential direct extensions of the work presented here. There is, of course, the possibility of further refining these factorizations, seeing if it is possible to remove some more of the remaining terms at higher orders and/or obtain an improvement of the convergence similar to that of the exponential resummation (particularly for the modes with large $m$). Even more enticing is the possibility of factorizing (or otherwise simplifying) many other quantities calculated using the MST formalism in a similar way. For instance, one might consider the phase of the
gravitational waveforms (also calculated to $22$PN by Fujita~\cite{Fujita22PN, Fujita22PN_web} for the Schwarzschild circular orbit case in the form of the Damour, Iyer, and Nagar~\cite{DIN} phase correction).

Additionally, it should also be possible to generalize the factorizations already developed here to apply to the gravitational wave flux at infinity from a particle in a circular, equatorial orbit in Kerr fairly straightforwardly, since the MST formalism experiences relatively minor changes in going from Schwarzschild to Kerr: This flux has recently been calculated to
$10$PN by Fujita~\cite{Fujita_prep} completely analytically (substantially improving upon the previous $4$PN calculation by Tagoshi \emph{et al.}~\cite{TSTS}), and has also been calculated numerically (with analytic forms determined for some coefficients) to $20$PN by Shah~\cite{Shah}. Similarly, it should likely be possible to develop a similar simplifying factorization for the horizon-absorbed flux (for Kerr or
Schwarzschild) for circular, equatorial orbits. This flux has recently been calculated to $8$PN by Fujita~\cite{Fujita_prep} completely analytically (again significantly improving upon the previous $4$PN accurate calculation by Tagoshi, Mano, and Takasugi~\cite{TMT}), and has also been calculated numerically (again with analytic forms determined for some coefficients) to $20$PN by Shah~\cite{Shah}. 

If these simplifications prove successful, then the natural extension would be to continue to generic (eccentric, inclined) orbits in Kerr, as discussed in Shah~\cite{Shah}, though here the expansions are not yet known to very high order (as reviewed in~\cite{ST}), so the factorization might be used to
assist in determining even relatively low-order coefficients analytically from numerical results, and developed in tandem with that calculation. Indeed, a desire to understand the structure of these expansions to aid in the analytic understanding of the numerical values of the coefficients was the initial motivation for the present research. As discussed by Bini and Damour~\cite{BD3}, knowing some of the form of a coefficient can greatly aid the determination of an analytic
form from a high-accuracy numerical result.

It might also be possible to perform a
similar simplifying factorization for Detweiler's redshift observable and the spin precession frequency, which have recently been calculated to $8.5$PN by Bini and Damour~\cite{BD3, BD4} completely analytically. The redshift observable has also been calculated numerically to $10.5$PN by Shah, Friedman, and Whiting~\cite{SFW}, with analytic forms determined for some coefficients.

Future work will also examine the convergence of the different factorizations and resummations of the different modes more systematically, and explore how the modes' convergence relates to the convergence of the PN series for $\nu$. Such work will also likely consider unstable orbits inside the ISCO: As discussed in~\cite{DIN,BNZ}, it is interesting to consider the convergence of the flux between the light ring and ISCO as a further test of the accuracy of various approximants, particularly since the flux there can be used to aid the modeling of the plunge phase of an extreme mass-ratio inspiral in the effective one-body (EOB) framework. In this case, it would likely be worthwhile to solve for $\nu$ numerically, as in Fujita and Tagoshi~\cite{FT2004,FT2005}. It would also be interesting to see how well the exponential resummation or factorizations such as those introduced here aid in increasing the agreement of PN gravitational wave amplitudes with numerical relativity results in the comparable-mass case. (See, e.g.,~\cite{Boyleetal} for such a comparison  for the dominant quadrupolar mode; \cite{Panetal,DNB} perform such a comparison for the higher modes using various EOB results.)

%%%%%%%%%%%
\acknowledgments
%%%%%%%%%%%

We thank Sebastiano Bernuzzi, Bernd Br{\"u}gmann, David Hilditch, and Georgios Loukes-Gerakopoulos for useful comments, and Tim Dietrich for a careful reading of the paper. We also owe special appreciation to Abhay G.\ Shah for calculating the PN expansions of $\nu$. This work was supported by the DFG SFB/Transregio 7.

%%%%%%%%%%

\bibliography{SFtoexactPN}

\end{document}